%% file: sample-sigconf.tex
\documentclass[10pt,conference]{IEEEtran}
\IEEEoverridecommandlockouts

\usepackage{cite}
\usepackage{amsmath,amssymb,amsfonts}
\usepackage{graphicx}
\usepackage{textcomp}
\usepackage{xcolor}
\def\BibTeX{{\rm B\kern-.05em{\sc i\kern-.025em b}\kern-.08em
    T\kern-.1667em\lower.7ex\hbox{E}\kern-.125emX}}

\def\BibTeX{{\rm B\kern-.05em{\sc i\kern-.025em b}\kern-.08emT\kern-.1667em\lower.7ex\hbox{E}\kern-.125emX}}

\newtheorem{definition}{Definition}
\newtheorem{proof}{Proof}

\DeclareMathOperator*{\argmin}{arg\,min}

\usepackage{lmodern}    
\usepackage{algorithm}
\usepackage{algorithmicx}
\usepackage{algpseudocode}
\usepackage{subcaption}
\usepackage{url}
\usepackage{framed}
\usepackage{adjustbox}
\begin{document}

%
\title{RobOT: Robustness-Oriented Testing for Deep Learning Systems}

\makeatletter
\newcommand{\linebreakand}{%
  \end{@IEEEauthorhalign}
  \hfill\mbox{}\par
  \mbox{}\hfill\begin{@IEEEauthorhalign}
}
\makeatother

\author{\IEEEauthorblockN{1\textsuperscript{st} Jingyi Wang}
\IEEEauthorblockA{
\textit{Zhejiang University}\\
wangjyee@zju.edu.cn}
\and
\IEEEauthorblockN{2\textsuperscript{nd} Jialuo Chen}
\IEEEauthorblockA{
\textit{Zhejiang University}\\
chenjialuo@zju.edu.cn}
\and
\IEEEauthorblockN{3\textsuperscript{rd} Youcheng Sun}
\IEEEauthorblockA{
\textit{Queen's University Belfast}\\
youcheng.sun@qub.ac.uk}
\and
\IEEEauthorblockN{4\textsuperscript{th} Xingjun Ma}
\IEEEauthorblockA{
\textit{Deakin University}\\
daniel.ma@deakin.edu.au}
\linebreakand
\IEEEauthorblockN{5\textsuperscript{th} Dongxia Wang}\thanks{* Dongxia Wang is the corresponding author.}
\IEEEauthorblockA{
\textit{Zhejiang University}\\
dxwang@zju.edu.cn}
\and
\IEEEauthorblockN{6\textsuperscript{th} Jun Sun}
\IEEEauthorblockA{
\textit{Singapore Management University}\\
junsun@smu.edu.sg}
\and
\IEEEauthorblockN{7\textsuperscript{th} Peng Cheng}
\IEEEauthorblockA{
\textit{Zhejiang University}\\
lunarheart@zju.edu.cn}
}

\maketitle

%
\begin{abstract}
Recently, there has been a significant growth of interest in applying software engineering techniques for the quality assurance of deep learning (DL) systems. One popular direction is deep learning testing, where adversarial examples (a.k.a.~bugs) of DL systems are found either by fuzzing or guided search with the help of certain testing metrics. 
However, recent studies have revealed that the commonly used neuron coverage metrics by existing DL testing approaches are not correlated to model robustness. It is also not an effective measurement on the confidence of the model robustness after testing. 
In this work, we address this gap by proposing a novel testing framework called \emph{Rob}ustness-\emph{O}riented \emph{T}esting (RobOT). 
A key part of RobOT is a quantitative measurement on 1) the value of each test case in improving model robustness (often via retraining), and 2) the convergence quality of the model robustness improvement. 
RobOT utilizes the proposed metric to automatically generate test cases valuable for improving model robustness. The proposed metric is also a strong indicator on how well robustness improvement has converged through testing.
Experiments on multiple benchmark datasets confirm the effectiveness and efficiency of RobOT in improving DL model robustness,
with 67.02\% increase on the adversarial robustness that is 50.65\% higher than the state-of-the-art work DeepGini.

\end{abstract}

\input{Introduction}
\input{Background}
\input{Methodology}
\input{Experiment}
\input{Related_Work}

\input{Conclusion}

\section*{Acknowledgments}
This work was supported by the National Key R\&D Program of China (Grant No. 2020YFB2010900). This work was also supported by the NSFC Program (Grant No. 62061130220, 61833015 and 62088101), the Guangdong Science and Technology Department (Grant No. 2018B010107004) and the National Research Foundation, Singapore under its AI Singapore Programme (AISG Award No.: AISG-RP-2019-012).

\bibliographystyle{plain}
\bibliography{convergence.bib}



\end{document}

%% file: Introduction.tex
\section{Introduction}

Deep learning (DL) \cite{dl} has been the core driving force behind the unprecedented breakthroughs in solving many challenging real-world problems such as object recognition \cite{cv} and natural language processing \cite{nlp}.
Despite the success, deep learning systems are known to be vulnerable to adversarial examples (or attacks), which are slightly perturbed inputs that are imperceptibly different from normal inputs to human observers but can easily fool state-of-the-art DL systems into making incorrect decisions \cite{fgsm,cw,ma2018characterizing,jiang2019black,wu2020skip}. 
This not only compromises the reliability and robustness of DL systems, but also raises security concerns on their deployment in safety-critical applications such as face recognition \cite{face}, malware detection \cite{md}, medical diagnosis \cite{finlayson2019adversarial,ma2020understanding} and autonomous driving \cite{av,duan2020adversarial}.

Noticeable efforts have been made in the software engineering community to mitigate the threats of adversarial examples and to improve the robustness of DL systems in the presence of adversarial examples \cite{deeppoly,deepxplore,wang2019adversarial}. Among them, formal verification aims to prove that no adversarial examples exist in the neighborhood of a given input. Substantial progress has been made using approaches like abstract interpretation \cite{deeppoly,yang2020improving} and reachability analysis \cite{star}. However, formal verification techniques are in general expensive and only scale to limited model structures and properties (e.g., local robustness \cite{huang2017safety}). 

Another popular line of work is deep learning testing, which aims to generate test cases that can expose the vulnerabilities of DL models. The test cases can then be used to improve the model robustness by retraining the model, \emph{however, this should not be taken as granted}, as recent studies have shown that test cases generated based on existing testing metrics have limited correlation to model robustness and robustness improvement after retraining~\cite{limit,meaningless}. In this work, we highlight and tackle the problem of effectively generating test cases for improving the adversarial robustness of DL models. 

\begin{figure*}[t]
\centering
\includegraphics[width=0.65\textwidth]{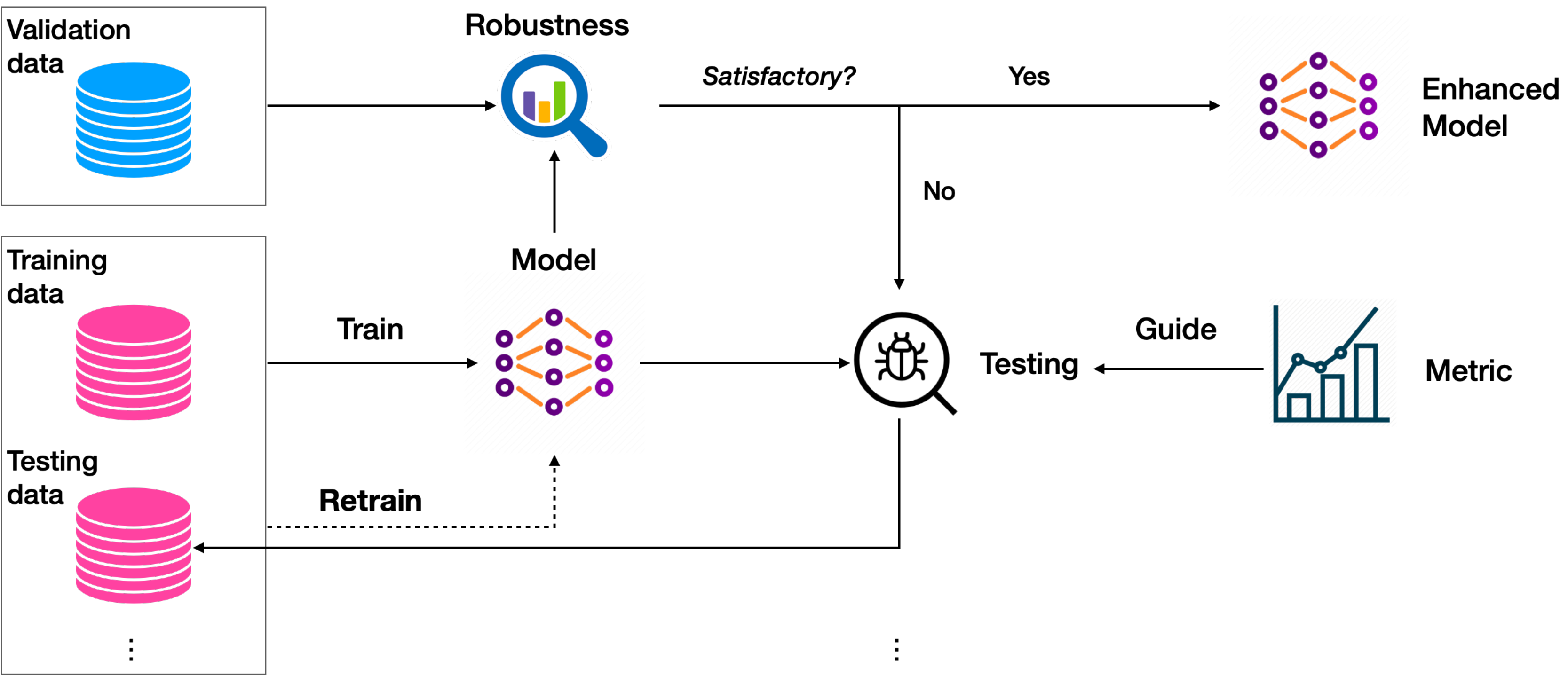}
\caption{Overview of RobOT testing framework.}
\label{fig:frame}
\end{figure*}


There are two key elements when it comes to testing DL systems. The first element is the testing metric used to evaluate the quality of a test case or a test suite. Multiple testing metrics, including neuron coverage~\cite{deepxplore}, multi-granularity neuron coverage~\cite{deepgauge} and surprise adequacy~\cite{surprise}, have been proposed. The common idea is to explore as much diversity as possible of a certain subspace defined based on different abstraction levels, e.g., neuron activation~\cite{deepxplore}, neuron activation pattern~\cite{deepgauge}, neuron activation conditions~\cite{concolic}, and neuron activation vector~\cite{surprise}. The second key element is the method adopted for test case generation, which is often done by manipulating a given seed input with the guidance of the testing metric. Existing test case generation techniques such as DeepXplore~\cite{deepxplore}, DeepConcolic~\cite{concolic}, DeepHunter~\cite{deephunter} and ADAPT~\cite{adapt} are mostly designed to improve the neuron coverage metrics of the test cases.
While existing testing approaches are helpful in exposing vulnerabilities of DL systems to some extent, recent studies have found that neuron coverage metrics are not useful for improving model robustness~\cite{limit,misleading,meaningless}. As a consequence, unlike in the case of traditional program testing (where the program is surely improved after fixing bugs revealed through testing), one may not improve the robustness of the DL system after testing.

In this work, we address the above-mentioned limitations of existing DL testing approaches by proposing a novel DL testing framework called RobOT (i.e., \textit{Rob}ustness-\textit{O}riented \textit{T}esting), which integrates the DL (re)training with the testing. 
As illustrated in Fig.~\ref{fig:frame}, RobOT distinguishes itself from existing neuron coverage guided testing works in the following important aspects. First, RobOT is robustness-oriented. RobOT takes a user-defined requirement on the model robustness as input and integrates the retraining process into the testing pipeline. RobOT iteratively improves the model robustness by generating test cases based on a testing metric and retraining the model. Second, in RobOT, we propose a novel set of lightweight metrics that are strongly correlated with model robustness. The metrics can quantitatively measure the relevance of each test case for model retraining, and are designed to favor test cases that can significantly improve model robustness, which is in contrast to existing coverage metrics that have little correlation with model robustness. Furthermore, the proposed metrics can in turn provide strong evidence on the model robustness after testing. 
The output of RobOT is an enhanced model that satisfies the robustness requirement.

In a nutshell, we make the following contributions.

\begin{itemize}
\item We propose a robustness-oriented testing (RobOT) framework for DL systems. 
RobOT provides an end-to-end solution for improving the robustness of DL systems against adversarial examples.

\item We propose a new set of lightweight testing metrics that quantify the importance of each test case with respect to the model's robustness, which are shown to be stronger indicators of the model's robustness than existing metrics.

\item We implement in RobOT, a set of fuzzing strategies guided by the proposed metrics to automatically generate high-quality test cases for improving the model robustness.

\end{itemize}

RobOT is publicly available as an open-source self-contained toolkit \cite{code}.
Experiments on four benchmark datasets confirm the effectiveness of RobOT in improving model robustness. Specifically, RobOT achieves 50.65\% more robustness improvement on average compared to state-of-the-art work DeepGini~\cite{deepgini}.


%% file: Background.tex
\section{Background}
\label{sec:back}


\subsection{Deep Neural Networks}

\begin{figure}[t]
\centering
\includegraphics[width=0.35\textwidth]{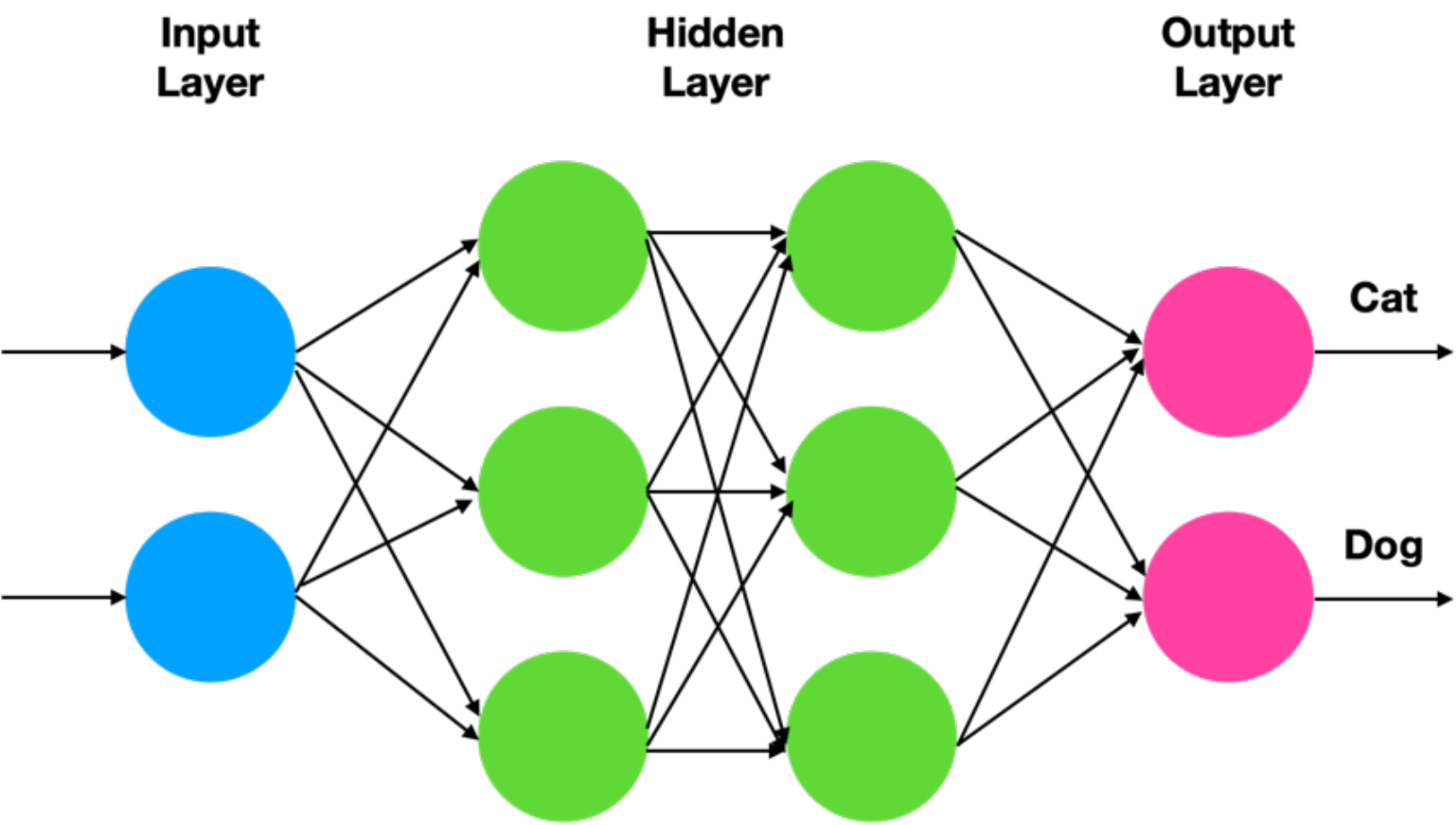}
\caption{An example DNN to predict cat or dog.}
\label{fig:dnn}
\end{figure}

In this work, we focus on deep learning models, e.g., deep neural networks (DNNs) for classification. 
We introduce a conceptual deep neural network (DNN) as an example in Fig.~\ref{fig:dnn} for simplicity and remark that our approach is applicable for state-of-the-art DNNs in our experiments like ResNet \cite{resnet}, VGG \cite{vgg}, etc.

\paragraph*{DNN}
\label{subsec:DNN}
A DNN classifier is a function $f:X\to Y$, which maps an input $x\in X$ (often preprocessed into a vector) into a label in $y\in Y$. As shown in Fig.~\ref{fig:dnn}, a DNN $f$ often contains an input layer, multiple hidden layers and an output layer. We use $\theta$ to denote the parameters of $f$ which assigns weights to each connected edge between neurons. 
Given an input $x$, we can obtain the output of each neuron on $x$, i.e., $f(x,ne)$, by calculating the weighted sum of the outputs of all the neurons in its previous layer and then applying an activation function (e.g., Sigmoid, hyperbolic tangent (tanh), or rectified linear unit (relu)) $\phi$. Given a dataset $D=\{(x_i,y_i)\}_{i=1}^n$, a DNN is often trained by solving the following optimization problem:
\begin{equation}
\min_\theta\frac{1}{n}\sum_{i=1}^n \mathcal{J}(f_\theta(x_i),y_i)
\end{equation}
, where $\mathcal{J}$ is a loss function which calculates a loss by comparing the model output $f_\theta(x_i)$ with the ground-truth label $y_i$. The most commonly used loss function for multi-class classification tasks is the categorical cross-entropy. The DNN is then trained by computing the gradient w.r.t. the loss for each sample in $D$ and updating $\theta$ accordingly.

\subsection{Deep Learning Testing} Most existing deep learning testing works are based on neuron coverage \cite{deepxplore} or its variants \cite{deepgauge}. Simply speaking, a neuron $ne$ is covered if there exists at least one test case $x$ where $f(x,ne)$ is larger than a threshold and thus been activated. We omit the details of other variants and briefly introduce the following testing methods as representatives. We also provide pointers for more details.

\vspace{1mm}
\noindent\textbf{DeepXplore} \cite{deepxplore} is the first testing work for DNN. DeepXplore proposed the first testing metric, i.e., neuron coverage and a differential testing framework to generate test cases to improve the neuron coverage.

\vspace{1mm}
\noindent\textbf{DeepHunter} \cite{deephunter} is a fuzzing framework which randomly selects seeds to fuzz guided by multi-granularity neuron coverage metrics defined in \cite{deepgauge}.

\vspace{1mm}
\noindent\textbf{ADAPT} \cite{adapt} is another recent work which adopts multiple adaptive strategies to generate test cases which could improve the multi-granularity neuron coverage metrics defined in \cite{deepgauge}.

\vspace{1mm}
\noindent\textbf{Adversarial Attacks} Beside the above testing methods, traditional adversarial attacks like FGSM \cite{fgsm}, JSMA \cite{jsma}, C\&W \cite{cw} and PGD \cite{pgd} attacks are also used to generate test cases in multiple works.

\subsection{Problem definition} 
Unlike existing coverage guided testing works, our goal is to design a robustness-oriented testing framework to improve the DL model robustness by testing. Two key problems are to be answered: 1) how can we design testing metrics which are strongly correlated with model robustness? 2) how can we automatically generate test cases favoring the proposed testing metrics?

%% file: Methodology.tex
\section{The RobOT Framework}
\label{sec:meth}


In this section, we present RobOT, a novel robustness-oriented framework for testing and re-training DL systems.
The overall framework of RobOT is shown in Figure~\ref{fig:frame}. We assume that \emph{a requirement on the model robustness} (Section \ref{sec:robustness}) is provided in prior for quality assurance
purpose. Note that the requirement is likely application-specific, i.e., different applications may have different requirements on the level of robustness.

RobOT integrates the DL (re)training into the testing framework. 
It starts from the initial training dataset $D_0$, and trains an initial DNN $f_0$ in the standard way. 
Then, it applies a fuzzing algorithm (see Section~\ref{sec:test-gen}) which is guided by our proposed testing metrics (see Section~\ref{sec:metrics})
to generate a new set of test cases $D_{t}$, for retraining the model $f_0$ to improve its adversarial robustness. 
The retraining step distinguishes RobOT from existing DL testing works and it places a specific requirement on how the test cases in $D_{t}$ are generated and selected, i.e., the test cases must be helpful in improving $f_0$'s robustness after retraining. We discuss how the test cases are generated in the rest of this section.

RobOT iteratively generates the test suite $D_{t}$ and retrains the model $f_{n}$ at each iteration. Afterwards, it checks whether the robustness of the new model $f_n$ is satisfactory using an independent adversarial validation dataset $D_v$, subject to an acceptable degrade of the model's accuracy on normal/non-adversarial data. If the answer is yes, it terminates and outputs the final model $f_{n}$; otherwise, RobOT continues until the model robustness is satisfactory or a predefined testing budget is reached. In the following, we illustrate each component of RobOT in detail.



\subsection{DL Robustness: A Formal Definition}
\label{sec:robustness}
Although many DL testing works in the literature claim a \emph{potential} improvement on the DL model robustness by retraining using the test suite generated, such a conjecture is often not rigorously examined. This is partially due to the ambiguous definition of robustness. For instance, the evaluations of \cite{deepxplore,deeptest,concolic,deephunter} are based on accuracy, in particular empirical accuracy on the validation set~\cite{testing-survey}, rather than robustness. In RobOT, we focus on improving the model \emph{robustness} (without sacrificing accuracy significantly), and we begin with defining robustness.
\begin{definition} \textbf{\textit{Global Robustness (GR)}} Given an input region $R$, a DL model $f:R\to Y$ is $(\sigma,\epsilon)$-globally-robust iff $\forall x_1,x_2\in R, ||x_1-x_2||_p\le\sigma \Rightarrow\ ||f(x_1)-f(x_2)||\le\epsilon$. $\hfill \square$
\end{definition}

Global robustness is theoretically sound, and yet extremely challenging for testing or verification~\cite{katz2017towards}. 
To mitigate the complexity, multiple attempts have been made to constrain the robustness into local input space, such as Local Robustness~\cite{huang2017safety}, CLEVER~\cite{clever} and Lipschitz Constant~\cite{xu2012robustness}. These local versions of robustness are however not ideal either, i.e., they have been shown to have their own limitations~\cite{lc_lim,katz2017towards}. For instance, CLEVER relies on the extreme value theory, making it extremely costly to calculate. 

In RobOT, we adopt a practical empirical definition of robustness, which has been commonly used for model robustness evaluation in the machine learning literature
\cite{pgd,pgd-con,zhang2019theoretically,wang2019improving,carlini2019evaluating}.

\begin{definition} \label{def:empirical-robustness}\textbf{\textit{Empirical Robustness (ER)}} Given a DL model $f:X\to Y$ and a validation dataset $D_v$, we define its empirical robustness $\mu:(f,D_v,ATT)\to [0,1]$ as $\gamma$, where $ATT$ denotes a given type of adversarial attack and $\gamma$ is the accuracy of $f$ on the adversarial examples obtained by conducting $ATT$ on $\langle D_v,f\rangle$. $\hfill \square$
\end{definition}

Intuitively, Def. \ref{def:empirical-robustness} evaluates a model's robustness using its accuracy on the adversarial examples crafted from a validation set $D_v$. Such an empirical view of DL robustness is testing-friendly and it facilitates RobOT to efficiently compare the robustness of the models before and after testing and retraining.  Definition~\ref{def:empirical-robustness} is also practical, as it connects the DL robustness with many existing adversarial attacks (such as~\cite{fgsm,pgd,cw}) as a part of the definition.
In particular, for the evaluation of RobOT in Section \ref{sec:exp}, we use two popular attacks, i.e., FGSM \cite{fgsm} and PGD (Projected Gradient Descent) \cite{pgd} as $ATT$.   


\subsection{RobOT DL Testing: A General View}

\begin{figure}[t]
\centering
\includegraphics[width=0.45\textwidth]{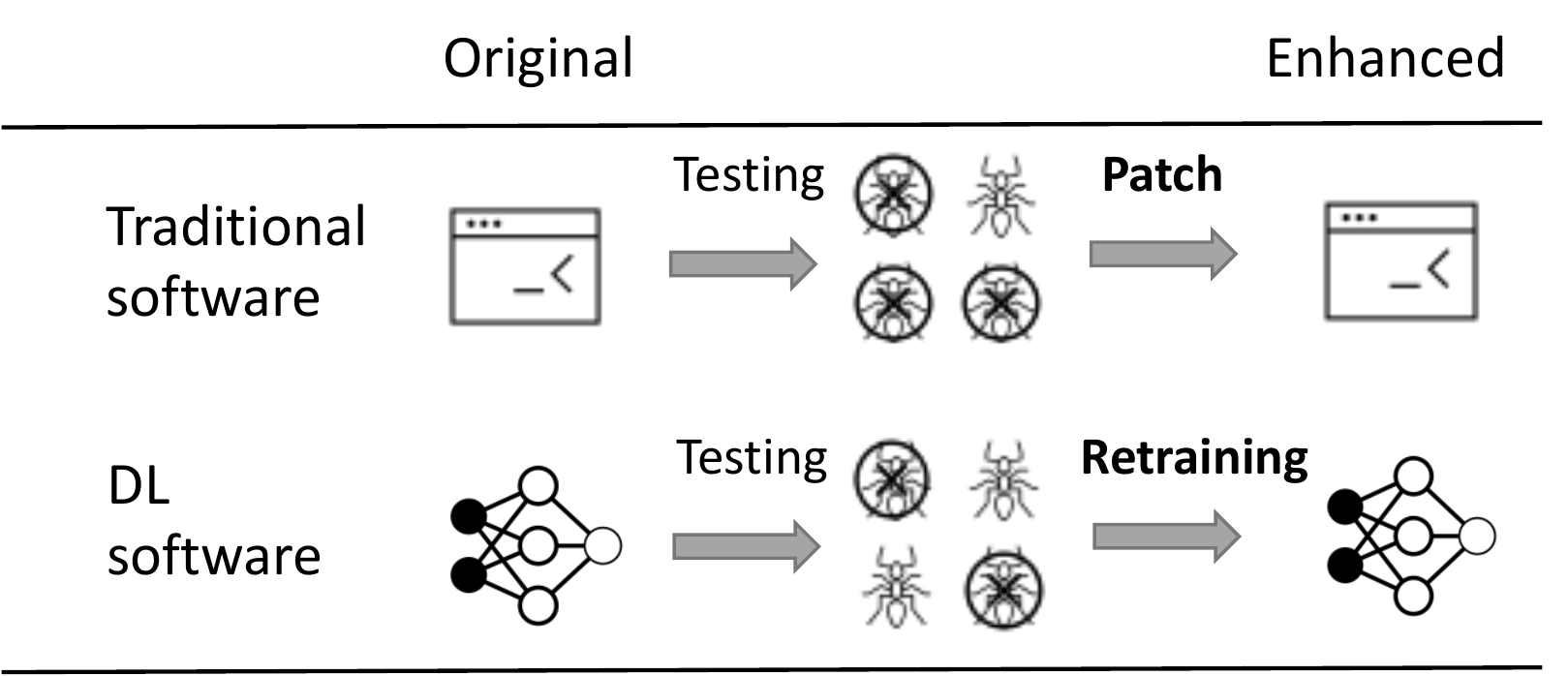}
\caption{Comparison between traditional and deep learning system quality assurance by testing.}
\label{fig:compar}
\end{figure}

We first compare and highlight the difference between testing traditional software and deep learning systems in Fig.~\ref{fig:compar}. While many testing methods (like random testing~\cite{randoop}, symbolic execution~\cite{klee}, concolic testing~\cite{wang2018towards} and fuzzing \cite{fuzz}) can be applied to identify vulnerabilities or bugs for both the traditional software and the DL systems, the workflow differs for the two after testing is done, i.e.,
the quality of traditional software is enhanced by patching the found bugs, whereas deep learning systems are improved via retraining.  
Arguably, the ultimate goal of testing is to improve the system's quality. Such improvement is guaranteed by patching bugs identified through testing in traditional software (assuming regression bugs are not frequent), i.e., the usefulness of a bug-revealing test for traditional software requires no justification. It is not obvious for DL systems, i.e., the usefulness of a test case can only be judged by taking into account the retraining step. Nevertheless, the retraining phase is largely overlooked so far in the deep learning testing literature.

Based on the Empirical Robustness definition in Def.~\ref{def:empirical-robustness}, in Alg.~\ref{alg:main}, we present the high level algorithmic design of RobOT for the workflow of DL testing in Figure \ref{fig:compar}. The initial trained model $f_0$ is given as an input in the algorithm and the testing and retraining iterations in RobOT are conducted within the main loop (Lines 2-6). The loop continues until the user-provided empirical robustness requirement is satisfied (Line 2). 

RobOT aims to bridge the gap between the DL testing and retraining. Let $T$ (Line 3) denote a fuzzing algorithm to generate test cases (guided by certain metrics). The objective of robustness-oriented testing is to improve the model robustness by testing. Formally, given a deep learning model $f$, the goal of RobOT at each iteration is to improve the following:
\begin{equation}
ER(\argmin_\theta\frac{1}{n}\sum_{i=1}^n\mathcal{J}_{(x_i,y_i)\in D\cup T(f,D)}(\theta,x_i,y_i)).
\end{equation} 
Intuitively, the testing metric should be designed in such a way that after retraining with the generated test cases, the model robustness is improved. This objective directly links the testing metric to the model robustness.

In the remaining of this section, we realize the method in Line 3 by answering the question: \emph{how should we design test metrics that are strongly correlated with the model robustness and how can we generate test cases guided by the proposed metrics?}



\begin{algorithm}[t]
\caption{RobOT($f_0,D,D_v,r,t$)}
\label{alg:main}
\begin{algorithmic}[1]
	\State $f=f_0$
	\While{$ER(f,D_v,t)<r$} 
		\State $D_t \leftarrow T(f, D)$ 
		\State $D \leftarrow D \cup D_t$
		\State Update $f$ by retraining the model with $D$
	\EndWhile
	\State \Return $f$
\end{algorithmic}
\end{algorithm}

\subsection{Robustness-Oriented Testing Metrics} 
\label{sec:metrics}
Our goal is to design testing metrics which are strongly correlated with model robustness. We note that there have been some efforts in the machine learning community to modify the standard training procedure in order to obtain a more robust model. For instance, the most effective and successful approach so far is robust training, which incorporates an adversary in the training process so that the trained model can be robust by minimizing the loss of adversarial examples in the first place \cite{pgd}:
\begin{equation}
\min_\theta\frac{1}{n}\sum_{i=1}^n\max_{||x'_i-x_i||_p\le\epsilon}\mathcal{J}(f(\theta,x'_i),y_i).
\end{equation}

At the heart of robust training is to identify a \emph{strong} (ideally worst-case) adversarial example $x'$ around\footnote{A $\epsilon-$ball defined according to a certain $L_p$ norm.} a normal example $x$ and train the model so that the loss on the strong adversarial example can be minimized. Robust training has shown encouraging results in training more robust models~\cite{pgd,pgd-con}. This inspires us to consider deep learning testing analogously in terms of how we generate test cases (around a normal example) and retrain the model with the test cases to improve the model robustness. The key implication is that when we design robustness-oriented testing metrics to guide testing, we should evaluate the usefulness of a test case from a loss-oriented perspective. 


Let $x_0$ be the seed for testing. We assume that a test case $x^t$ is generated in the neighborhood $\epsilon-$ball around $x_0$, i.e., $\{x|\ ||x-x_0||_p\le\epsilon\}$ using either a testing method or an adversarial attack. The main intuition is that a test case which induces a higher loss is a stronger adversarial example, which is consequently more helpful in training robust models~\cite{pgd}. Based on this intuition, we propose two levels of testing metrics on top of the loss as follows.

\paragraph{Zero-Order Loss (ZOL)} The first metric directly calculates the loss of a test case with respect to the DL model. Formally, given a test case $x^t$ (generated from seed $x$), a DL model $f$, the loss of $x^t$ on $f$ is defined as:
\begin{equation}
ZOL(x^t,f)=\mathcal{J}(f(\theta,x^t),y),
\end{equation}
where $y$ is the ground-truth label of $x$. For test cases generated from the same seed, we prefer test cases with higher loss, which are more helpful in improving the model robustness via retraining.

\paragraph{First-Order Loss (FOL)} The loss of generated test cases can be quite different for different seeds. In general, it is easier to generate test cases with high loss around seeds which unfortunately do not generalize well. Thus, ZOL is unable to measure the value of the test cases in a unified way. To address this problem, we propose a more fine-grained metric which could help us measure to what degree we have achieved the highest loss in the seed's neighborhood. The intuition is that, given a seed input, the loss around it often first increases and eventually converges if we follow the gradient direction to modify the seed~\cite{pgd}. Thus, a criteria which measures how well the loss converges can serve as the testing metric. A test case with better convergence quality corresponds to a higher loss than its neighbors. Next, we introduce First-Order Stationary Condition (FOSC) to provide a measurement on the loss convergence quality of the generated test cases. 

Formally, given a seed input $x_0$, its neighborhood area $\mathcal{X}=\{x|\ ||x-x_0||_p\le\epsilon\}$, and a test case $x^t$, the FOSC value of $x^t$ is calculated as:
\begin{equation}
\label{eq:fosc}
c(x^t)= \text{max}_{x\in \mathcal{X}}\langle x-x^t,\nabla_x f(\theta,x^t)\rangle.
\end{equation}

In~\cite{pgd-con}, it is proved that the above problem has the following closed form solution if we take the $\infty-$norm for $\mathcal{X}$.
\begin{equation}
\label{eq:fosc:sol1}
c(x^t)=\epsilon||\nabla_x f(\theta,x^t)||_1-\langle x^t-x_0,\nabla_x f(\theta,x^t)\rangle.
\end{equation}

However, many existing DL testing works are generating test cases from the $L_2$ norm neighborhood which makes the above closed-form solution for $L_\infty$ infeasible. We thus consider solving the formulation in Eq. \ref{eq:fosc} with $L_2$ norm and obtain the solution as follows:
\begin{equation}
\label{eq:fosc:sol2}
c(x^t)=\epsilon||\nabla_x f(\theta,x^t)||_2.
\end{equation}

\begin{proof}
According to Cauchy–Schwarz inequality:

\begin{align*}
|\langle x-x^t,\nabla_x f(\theta,x^t)\rangle|^2\le\\ 
\langle x-x^t,x-x^t\rangle \cdot \langle \nabla_x f(\theta,x^t),\nabla_x f(\theta,x^t)\rangle\le\\
\epsilon^2\cdot (||\nabla_x f(\theta,x^t)||_2)^2
\end{align*}

Since there must exist $x^t$ such that $x-x^t$ and $\nabla_x f(\theta,x^t)$ are in the same direction, we thus have:
\begin{align*}
\max|\langle x-x^t,\nabla_x f(\theta,x^t)\rangle|^2=\epsilon^2\cdot(||\nabla_x f(\theta,x^t)||_2)^2
\end{align*}
Thus,
\begin{align*}
\max|\langle x-x^t,\nabla_x f(\theta,x^t)\rangle|=\epsilon\cdot ||\nabla_x f(\theta,x^t)||_2
\end{align*}
\end{proof}

Note that FOSC (in both Eq.~\ref{eq:fosc:sol1} and Eq. \ref{eq:fosc:sol2}) is cheap to calculate, whose main cost is a one-time gradient computation (easy to obtain by all the DL frameworks). The FOSC value represents the first-order loss of a given test case. The loss of a test case converges and achieves the highest value if its FOSC value equals zero. Thus, a smaller FOSC value means a better convergence quality and a higher loss. 





\paragraph{Comparison with Neuron Coverage Metrics} 
Compared to neuron coverage metrics, our proposed loss based metrics have the following main differences. First, both ZOL and FOL are strongly correlated to the adversarial strength
of the generated test cases and the model robustness. Thus, our metrics can serve as strong indicators on the model's robustness after retraining. Meanwhile, our metrics are also able to measure the value of each test case in retraining, which helps us select valuable test cases from a large amount of test cases to reduce the retraining cost.


\subsection{FOL Guided Test Case Selection} In the following, we show the usefulness of the proposed metric through an important application, i.e., test case selection from a massive amount of test cases. Note that by default, we use the FOL metric hereafter due to the limitation of ZOL as described above. Test case selection is crucial for improving the model robustness with limited retraining budget. The key of test case selection is to quantitatively measure the value of each test case. So far this problem remains an open challenge. Prior work like DeepGini has proposed to calculate a Gini index of a test case from the model's output probability distribution~\cite{deepgini}. DeepGini's intuition is to favor those test cases with most uncertainty (e.g., a more flat distribution) under the current model's prediction. Compared to DeepGini, FOL contains fine-grained information at the loss level and is strongly correlated with model robustness. 

Given a set of test cases $D^t$, we introduce two strategies based on FOL to select a smaller set $D^s\subset D^t$ for retraining the model as follows. Let $D^t=[x_1,x_2,\cdots,x_m]$ be a ranked list in descending order by FOL value, i.e., $FOL(x_i)\ge FOL(x_{i+1})$ for $i\in[1,m-1]$.

\begin{algorithm}[t]
\caption{KM-ST($D^t,k,n$)}
\label{alg:kmst}

\begin{algorithmic}[1]
\State $D^s=\emptyset$
\State Let $max$ and $min$ be the maximum and minimum FOL value respectively
\State Equally divide range $[min,max]$ into $k$ sections $KR=[R_1,R_2,\cdots,R_k]$
\For{Each FOL range $r\in[R_1,R_2,\cdots,R_k]$}
	\State Randomly select $n/k$ samples $D^r$ from $D^t$ whose FOL values are in $r$
	\State $D^s=D^s\cup D^r$
\EndFor
	\State \Return $D^s$
\end{algorithmic}
\end{algorithm}

\paragraph{K-Multisection Strategy (KM-ST)} The idea of KM-ST is to uniformly sample the FOL space of $D^t$. Algo. \ref{alg:kmst} shows the details. Assume we need to select $n$ test cases from $D^t$. We equally divide the range of FOL into $k$ sections ($KR$) at line 3. 
Then for each range $r\in KR$, we randomly select the same number of test cases at line 5.

\paragraph{Bi-End Strategy (BE-ST)} The idea of BE-ST is to form $D^s$ by equally combining test cases with small and large FOSC values. This strategy mixes test cases of strong and weak adversarial strength, which is inspired by a recent work on improving standard robust training~\cite{khoury2019adversarial}. Given a ranked $D^t$, we can simply take an equal number of test cases from the two ends of the list to compose $D^s$.

\begin{figure*}[t]
\begin{subfigure}[b]{0.25\textwidth}
   \centering 
   \includegraphics[height=	1.45in]{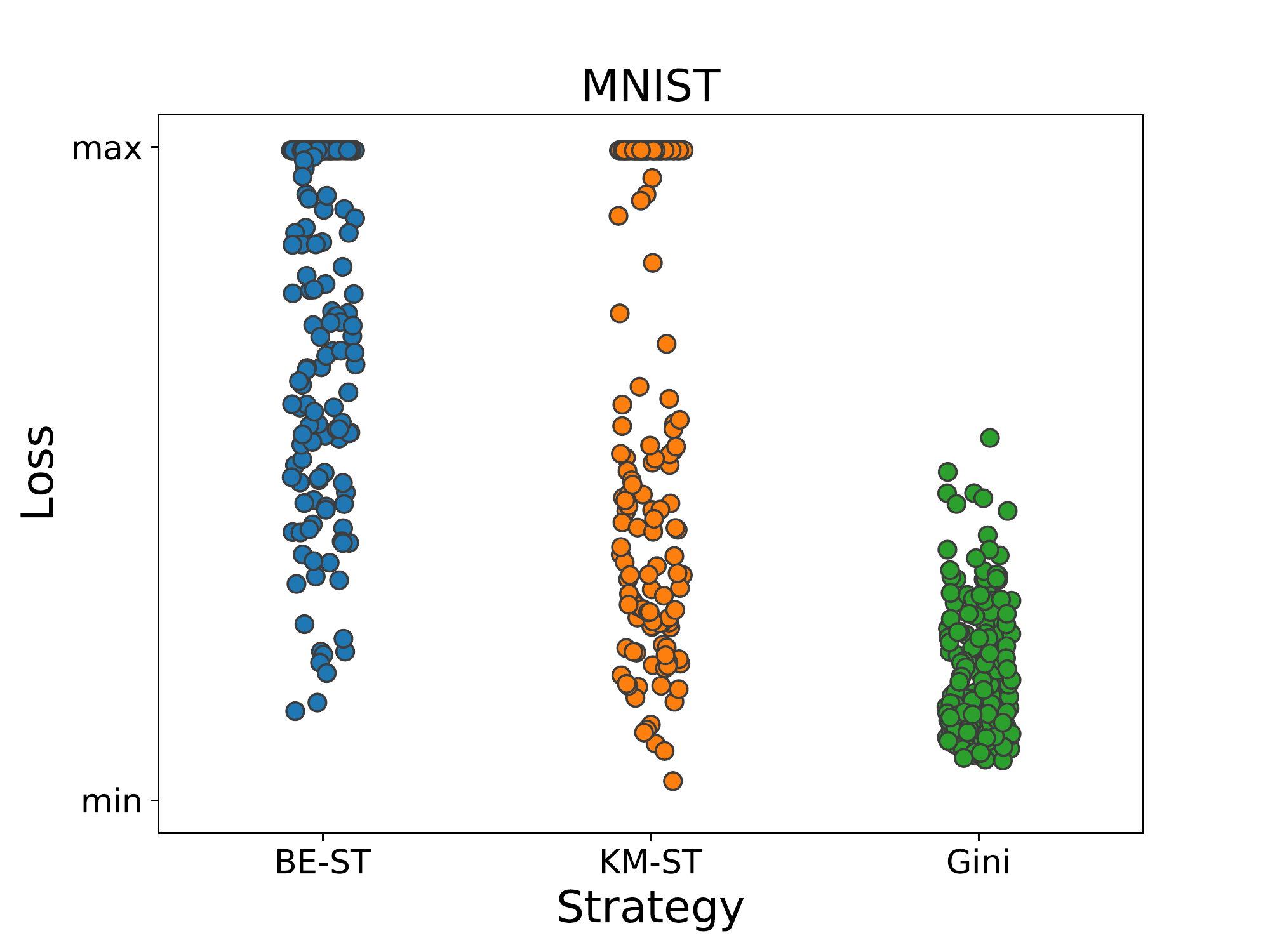}
   \label{fig:st:mnist}
\end{subfigure}%
\begin{subfigure}[b]{0.25\textwidth}
   \centering 
   \includegraphics[height=1.45in]{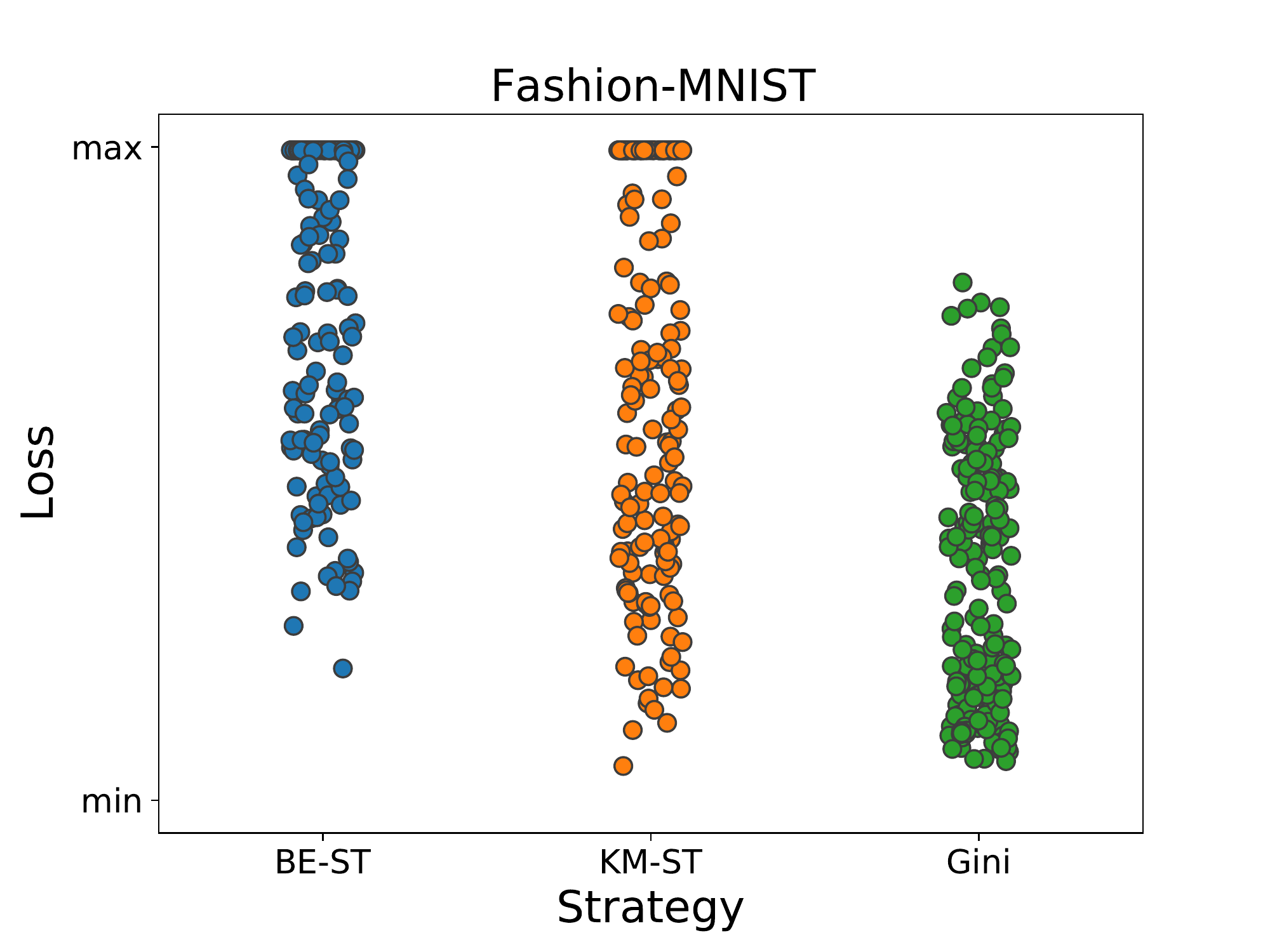}
   \label{fig:st:fashion}
\end{subfigure}%
\begin{subfigure}[b]{0.25\textwidth}
   \centering 
   \includegraphics[height=1.45in]{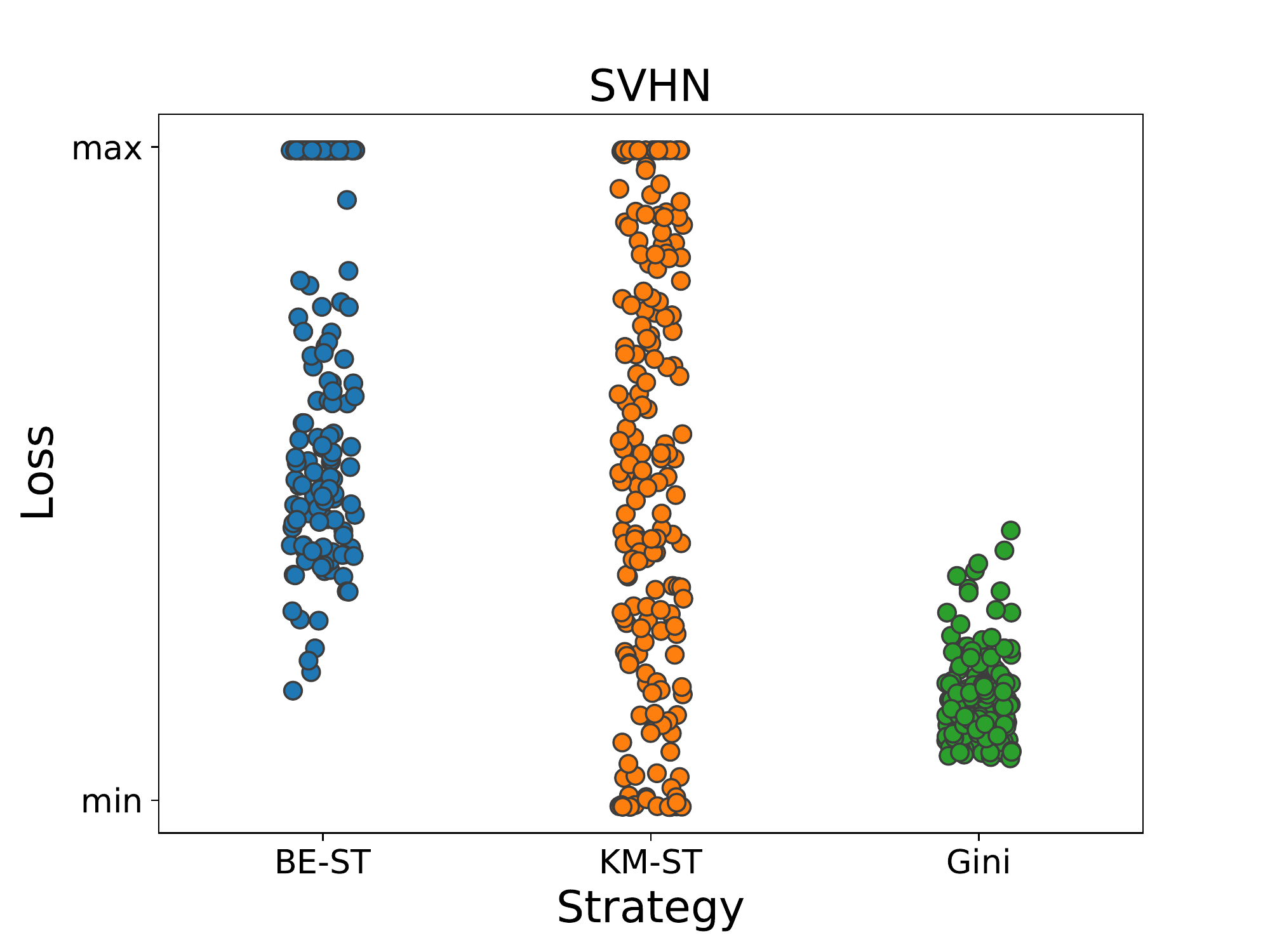}
   \label{fig:st:svhn}
\end{subfigure}%
\begin{subfigure}[b]{0.25\textwidth}
   \centering 
   \includegraphics[height=1.45in]{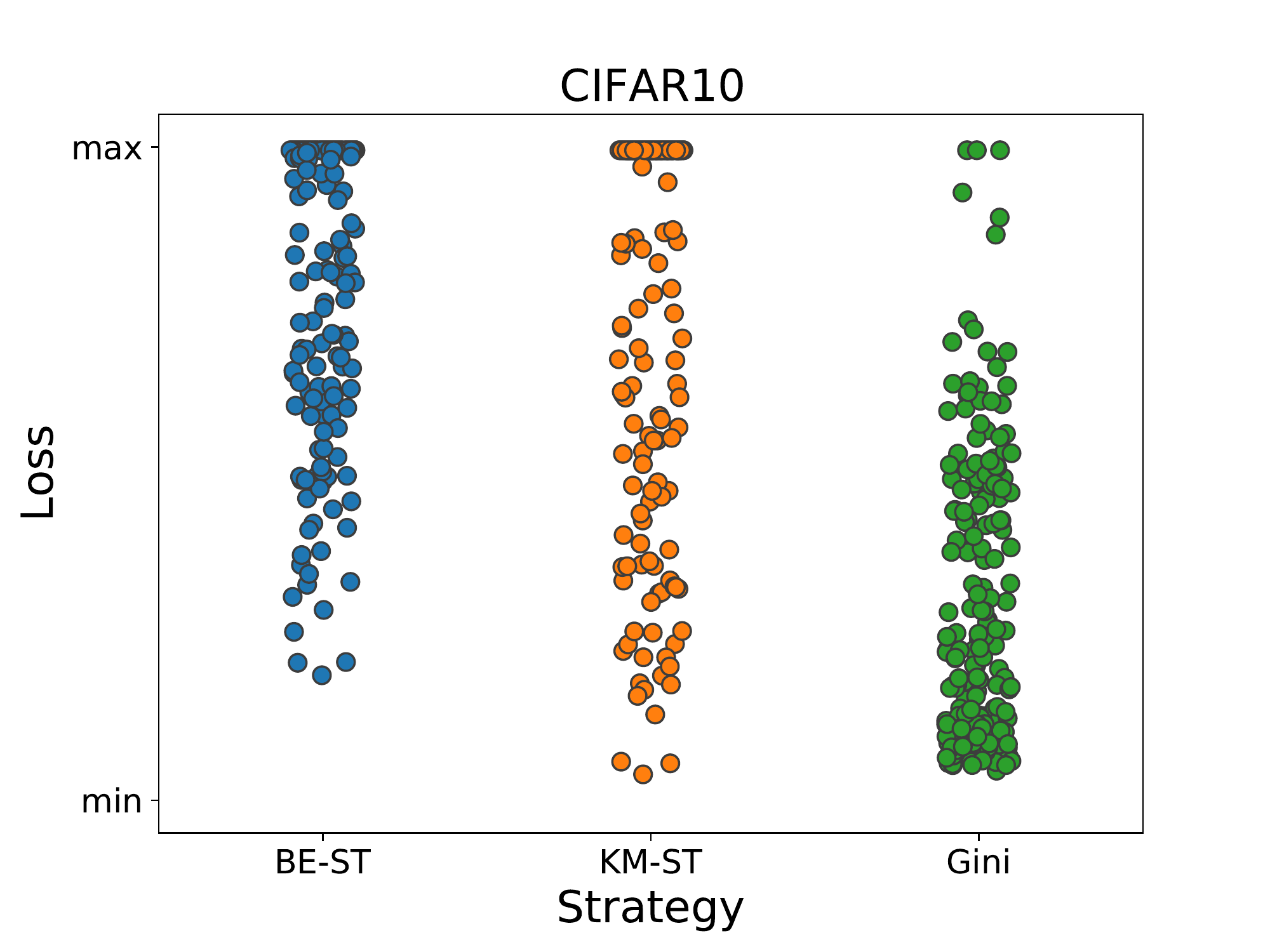}
   \label{fig:st:cifar}
\end{subfigure}%
   \caption{Loss of selected test cases for different datasets using different strategies.}
   \label{fig:loss:st}
\end{figure*}

Figure~\ref{fig:loss:st} shows the loss map of the selected test cases according to different strategies. We could observe that BE-ST prefers test cases of higher loss, KM-ST uniformly samples the loss space, while DeepGini often prefers test cases with lower loss.

\subsection{FOL Guided Fuzzing} 
\label{sec:test-gen}

Next, we introduce a simple yet efficient fuzzing strategy to generate test cases based on FOL. Note that since we have no prior knowledge of the FOL distribution, we are not able to design fuzzing strategy for KM-ST. Instead, we design a fuzzing algorithm for the BE-ST strategy. The idea is to greedily search for test cases in two directions, i.e., with both small or large FOL values.

Algo.~\ref{alg:fuzz} presents the details. The inputs include the model $f$, the list of seeds to fuzz $seeds\_list$, the fuzzing region $\epsilon$, the threshold on the small FOL value $\xi$, the number of labels to optimize $k$, a hyper-parameter $\lambda$ on how much we favor FOL during fuzzing and lastly the maximum number of iterations to fuzz for a seed $iters$. For each seed in the list, we maintain a list of seeds $s\_list$ at line 3. After obtaining a seed $x$ from $s\_list$ (line 5), we iteratively add perturbation on it from line 8 to line 28 in a way guided by FOL. We set the following objective for optimization (line 9).
\begin{equation}
\label{eq:obj}
obj=\sum_{i=2}^k P(c_i)-P(c_1)+\lambda \cdot FOL(x')
\end{equation}   
, where $c_i$ is the label with the $i^{\text{th}}$ largest softmax probability of $f$ ($c_1$ with the maximum), $P(c)$ is the softmax output of label $c$ and $k$ is a hyper-parameter. The idea is to guide perturbation towards changing the original label (i.e., generating an adversarial example) whilst increasing the FOL value. We then obtain the gradient of the objective (line 10) and calculate the perturbation based on the gradient by multiplying a learning rate and a randomized coefficient (0.5 to 1.5) to avoid duplicate perturbation (line 11). We run two kinds of checks to achieve the BE-ST strategy at line 15 and line 22 respectively. If the FOL value of the new sample after perturbation (x') is either increasing (line 15) or is smaller than a threshold (line 22), we add $x'$ to the seed list (line 17 and line 23). Furthermore, we add $x'$ to the fuzzing result if it satisfies the check and has a different label with the original seed $x$ (line 19 and line 25). Note that compared to neuron coverage guided fuzzing algorithms which need to profile and update neuron coverage information~\cite{adapt,deephunter}, our FOL guided fuzzing algorithm is much more lightweight, i.e., whose main cost is to calculate a gradient at each step. 

\begin{algorithm}[t]
\caption{FOL-Fuzz($f,seeds\_list,\epsilon,\xi,k,\lambda,iters$)}
\label{alg:fuzz}

\begin{algorithmic}[1]
\State Let $fuzz\_result=\emptyset$
\For{$seed\in seeds\_list$}
	\State Maintain a list $s\_list=[seed]$
	\While{$s\_list$ is not empty}
		\State Obtain a seed $x=s\_list.pop()$
		\State Obtain the label of the seed $c_1=f(x)$
		\State Let $x'=x$
		\For{$iter=0$ to $iters$}
			\State Set optimization objective $obj$ using Eq.~\ref{eq:obj}
			\State Obtain $grads = \frac{\nabla obj}{\nabla x'}$
			\State Obtain $perb = processing(grads)$
			\State Let $x’ = x' + perb$
			\State Let $c'= f(x’)$
			\State Let $dis = Dist(x’, x)$
			\If{$FOL(x')\ge FOL_m$ and $dis\le\epsilon$}
				\State $FOL_m=FOL(x')$
				\State $s\_list.append(x')$
				\If{$c'!=c_1$}
				\State $fuzz\_result.append(x')$
				\EndIf
			\EndIf
			\If{$FOL(x')<\xi$ and $dis\le\epsilon$}
				\State $s\_list.append(x')$
				\If{$c'!=c_1$}
				\State $fuzz\_result.append(x')$
				\EndIf
			\EndIf
		\EndFor

	\EndWhile
\EndFor
\State \Return $fuzz\_result$
\end{algorithmic}
\end{algorithm}

%% file: Experiment.tex
\section{Experimental Evaluation}
\label{sec:exp}

We have implemented RobOT as a self-contained tookit with about 4k lines of Python code. The source code and all the experiment details are available at \cite{code}. In the following, we evaluate RobOT through multiple experiments.

\subsection{Experiment Settings}

\paragraph{Datasets and Models} We adopt four widely used image classification benchmark datasets for the evaluation. We summarize the details of the datasets and models used in Tab. \ref{tb:ds}. 


\paragraph{Test Case Generation} We adopt two kinds of adversarial attacks and three kinds of coverage-guided testing approaches
to generate test cases for the evaluation in the following.

We summarize all the configurations of the test case generation algorithms in Tab.~\ref{tb:tg}.

\paragraph{Test Case Selection Baseline} We adopt the most recent work DeepGini \cite{deepgini} as the baseline of the test case selection strategy. DeepGini calculates a Gini index for each test case according to the output probability distribution of the model. A test case with larger Gini index is considered more valuable for improving model robustness.

\paragraph{Robustness Evaluation} We adopt Def. \ref{def:empirical-robustness} to empirically evaluate a model's robustness. In practice, we compose a validation set of adversarial examples $D_v$ for each dataset by combining the adversarial examples generated using both FGSM and PGD (10000 each). The attack parameters are the same with Tab. \ref{tb:tg}. We then evaluate a model's robustness by calculating its accuracy on $D_v$.

\begin{table}[]
\centering
\caption{Datasets and models.}
\begin{tabular}{|l|llll|}
\hline
Dataset       & Training & Testing & Model     & Accuracy \\ \hline
MNIST         & 60000    & 10000   & LeNet-5   & 99.02\%  \\
Fashion-MNIST & 60000    & 10000   & LeNet-5   & 90.70\%  \\
SVHN          & 73257    & 26032   & LeNet-5   & 88.84\%  \\
CIFAR10       & 50000    & 10000   & ResNet-20 & 90.39\%  \\
\hline
\end{tabular}
\label{tb:ds}
\end{table}

\begin{table}[]
\centering
\caption{Test case generation details.}
\scalebox{0.74}{
   \begin{tabular}{|l|l|llll|}
\hline
Testing Method & Parameter      & MNIST & SVHN   & Fashion-MNIST & CIFAR10 \\ \hline
FGSM           & Step size      & 0.3   & 0.03   & 0.03          & 0.01    \\
PGD            & Steps          & 10    & 10     & 10            & 10      \\
               & Step size      & 0.3/6 & 0.03/6 & 0.3/6        & 0.01/6  \\
DeepXplore     & Relu threshold & 0.5   & 0.5    & 0.5           & 0.5     \\
DLFuzz/ADAPT   & Time per seed  & 10 s    & 10 s     & 10 s            & 20 s      \\
               & Relu threshold & 0.5   & 0.5    & 0.5           & 0.5     \\

\hline
\end{tabular}
}

\label{tb:tg}
\end{table}

\subsection{Research Questions}

\vspace{1mm}
\noindent\textbf{RQ1: What is the correlation between our FOL metric and model robustness?} To answer this question, we first select three models with different robustness levels for each dataset. The first model (Model 1) is the original trained model. The second model (Model 2) is a robustness-enhanced model which is retrained\footnote{Retaining in this work takes 10 (40 for CIFAR10) additional epochs based on the original model.} by augmenting 5\% of the generated test cases and is more robust than Model 1. The third model (Model 3) is a robustness-enhanced model which is retrained by augmenting 10\% of the generated test cases and is most robust. Then, for each model, we conduct adversarial attacks to obtain a same number (10000 for FGSM and 10000 for PGD) of adversarial examples.

\begin{figure*}[t]
\begin{subfigure}[b]{0.25\textwidth}
   \centering 
   \includegraphics[height=1.45in]{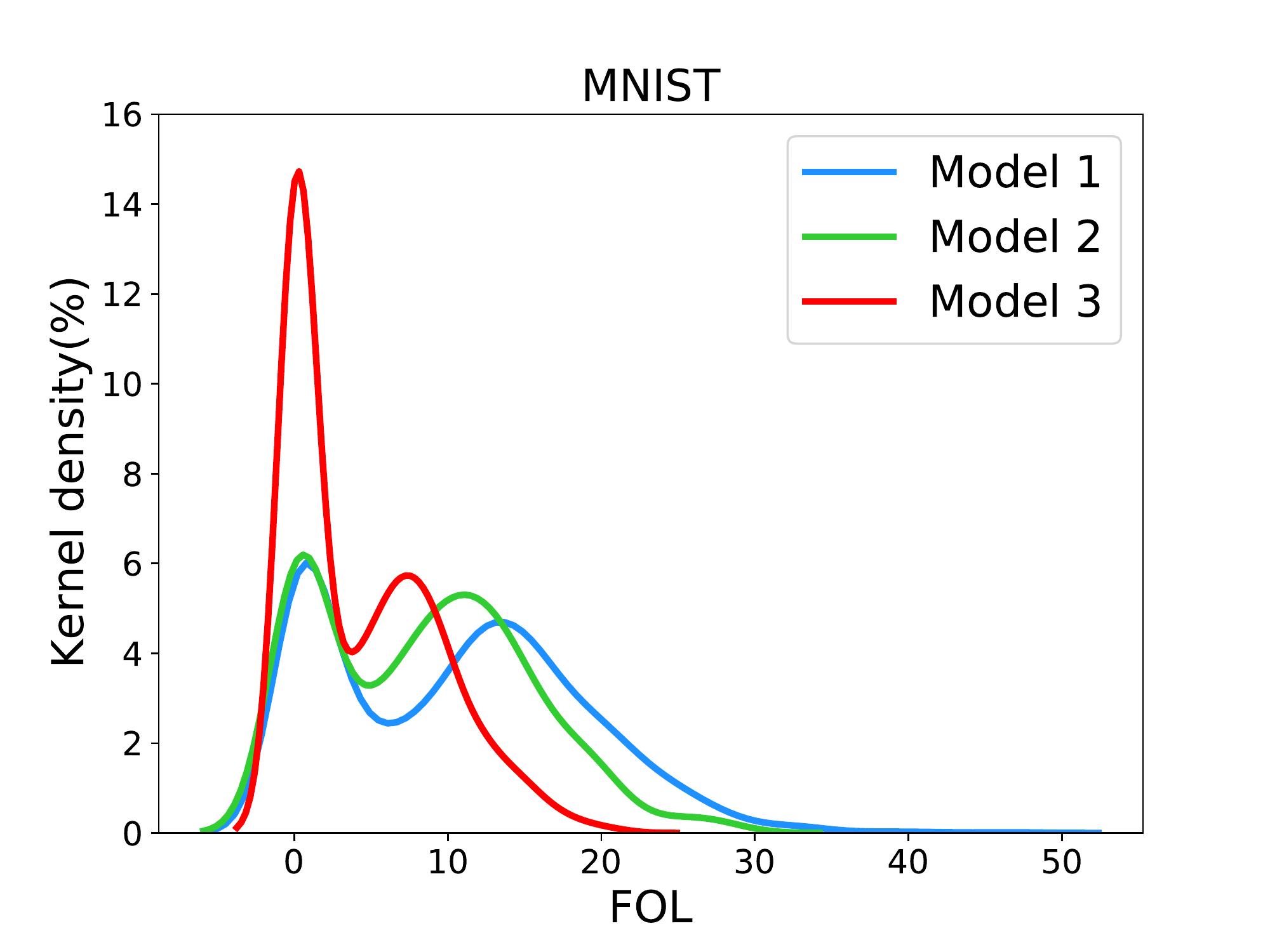}
   \label{fig:fol:mnist}
\end{subfigure}%
\begin{subfigure}[b]{0.25\textwidth}
   \centering 
   \includegraphics[height=1.45in]{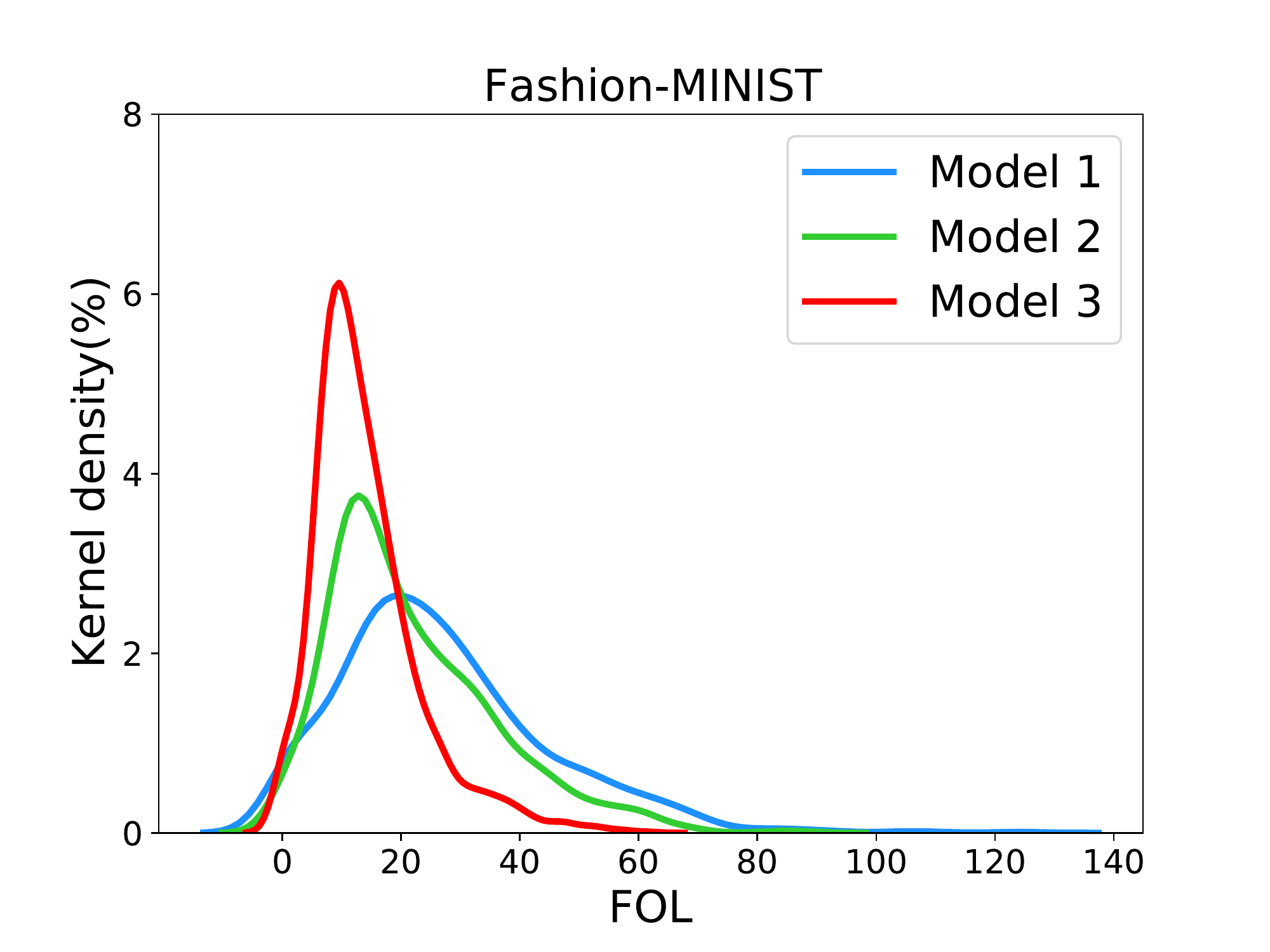}
   \label{fig:fol:fashion}
\end{subfigure}%
\begin{subfigure}[b]{0.25\textwidth}
   \centering 
   \includegraphics[height=1.45in]{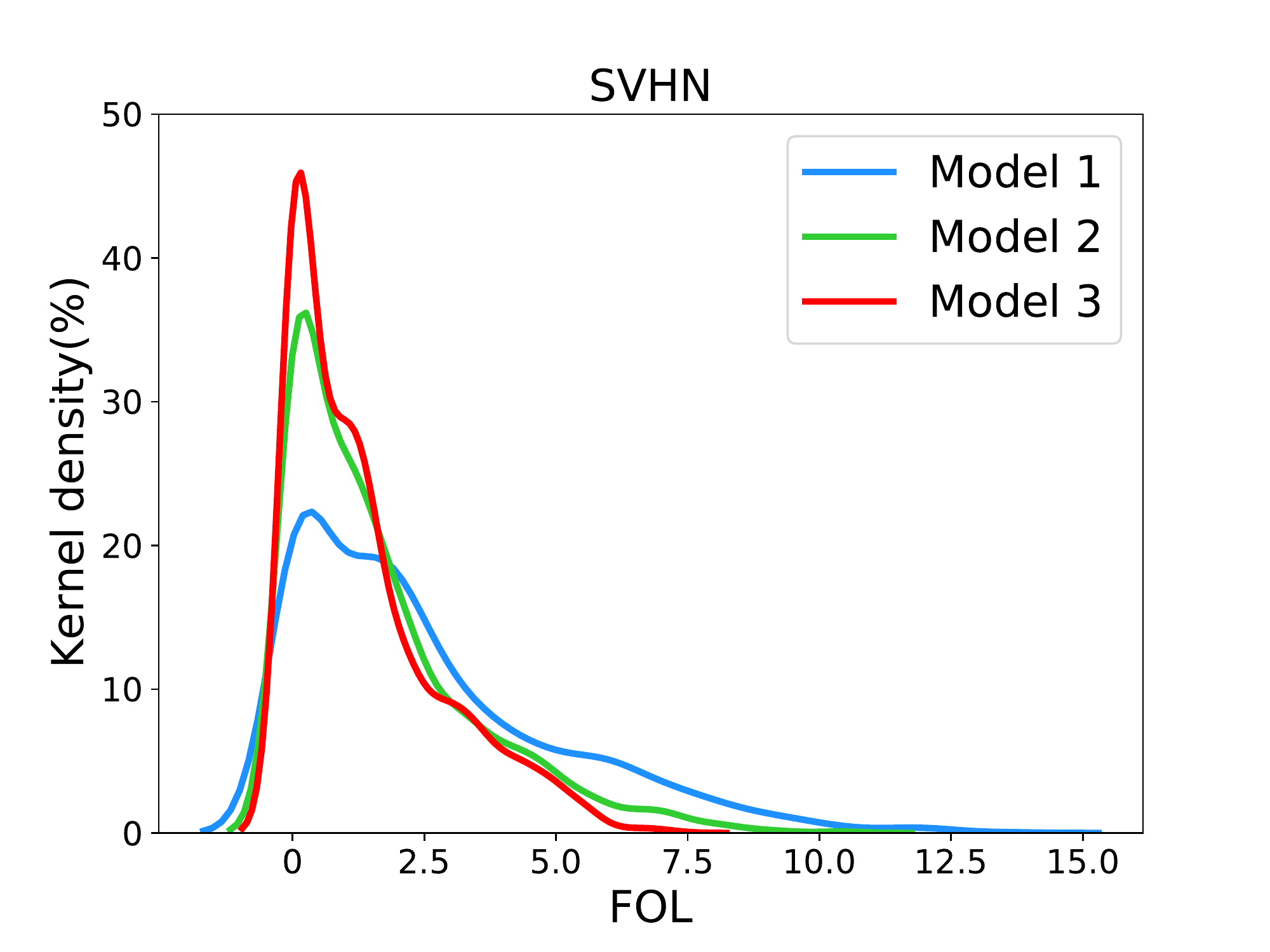}
   \label{fig:fol:svhn}
\end{subfigure}%
\begin{subfigure}[b]{0.25\textwidth}
   \centering 
   \includegraphics[height=1.45in]{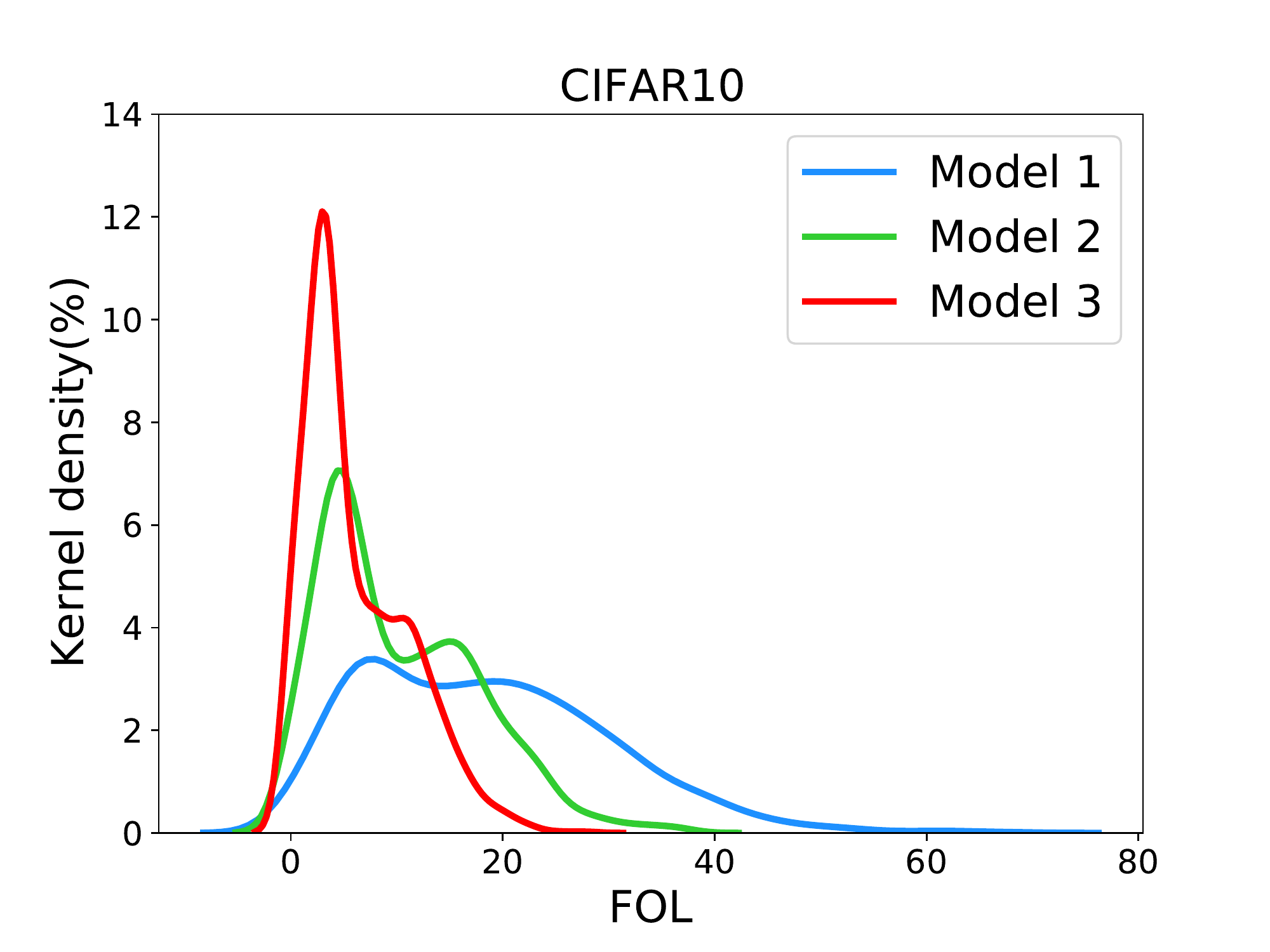}
   \label{fig:fol:cifar}
\end{subfigure}%
   \caption{FOL distribution of adversarial examples for models with different robustness.}
   \label{fig:fol}
\end{figure*}

We show the FOL distribution of the adversarial examples for different models in Fig. \ref{fig:fol}. We observe that there is a strong correlation between the FOL distribution of adversarial examples and the model robustness. Specifically, \emph{the adversarial examples of a more robust model have smaller FOL values.} This is clearly evidenced by Fig. \ref{fig:fol}, i.e., for every dataset, the probability density is intensively distributed around zero for Model 3 (the most robust model) while is steadily expanding to larger FOL values for Model 2 and Model 1 (with Model 1 larger than Model 2). The underlying reason is that a more robust model in general has a more \emph{flat} loss distribution and thus a smaller FOL value (since it is based on the loss gradient). 

\begin{figure}[t]
\centering 
\includegraphics[width=.33\textwidth]{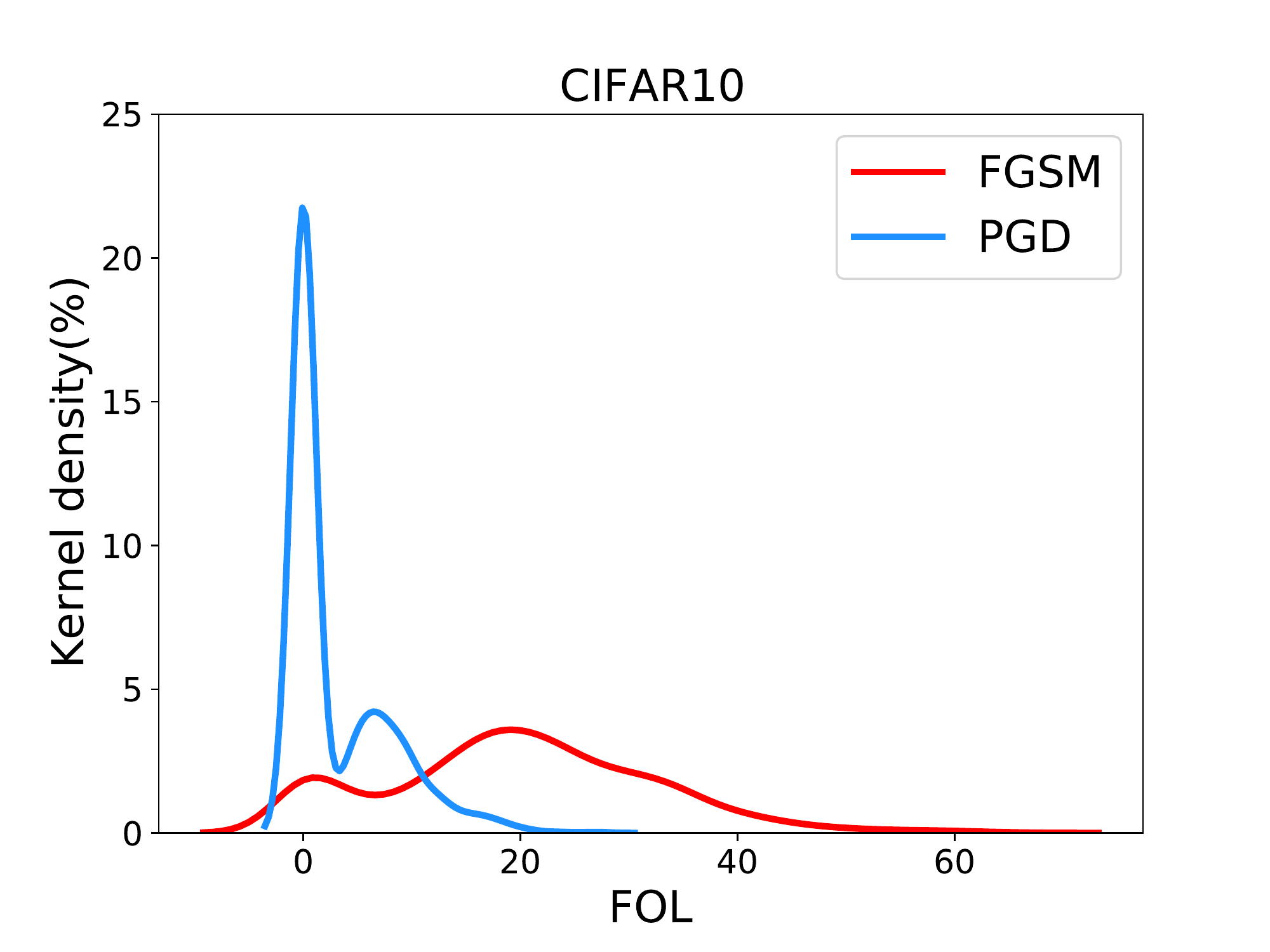}
\caption{FOL distribution of adversarial examples from FGSM and PGD for CIFAR10 model.} 
\label{fig:fol:com}
\end{figure}

In addition, we also observe that adversarial examples crafted by stronger attacks have smaller FOL values. Fig. \ref{fig:fol:com} shows the FOL distribution of adversarial examples from attacking the CIFAR10 model with FGSM and PGD respectively. We could observe that adversarial examples from PGD have significantly smaller FOL values than FGSM. The reason is that stronger attacks like PGD are generating adversarial examples that have better loss convergence quality and induce higher loss. 

We thus have the following answer to RQ1:
\begin{framed}
\noindent \emph{Answer to RQ1: FOL is strongly correlated with model robustness. A more robust model have smaller FOL values for adversarial examples.
}\end{framed}       




\vspace{1mm}
\noindent\textbf{RQ2: How effective is our FOL metric for test case selection?} To answer the question, we first generate a large set of test cases using different methods, and then adopt different test case selection strategies (i.e., BE-ST, KM-ST and DeepGini) to select a subset of test cases with the same size to retrain the model. A selection strategy is considered more effective if the retrained model with the selected test cases is more robust.

We distinguish two different kinds of test case generation algorithms which are both used in the literature, i.e., adversarial attacks and neuron coverage-guided algorithms, for more fine-grained analysis. For adversarial attacks, we adopt FGSM (weak) and PGD (strong) attacks to generate a combined set of test cases. For DeepXplore, DLFuzz and ADAPT, we generate a set of test cases for each of them. The parameters used are consistent with Tab. \ref{tb:tg}. For each set of test cases, we use BE-ST, KM-ST and DeepGini strategy respectively to select x (ranging from 1 to 10) percent of them to obtain a retrained model and evaluate its model robustness.

\begin{figure*}[t]
\begin{subfigure}[b]{0.25\textwidth}
   \centering 
   \includegraphics[height=1.45in]{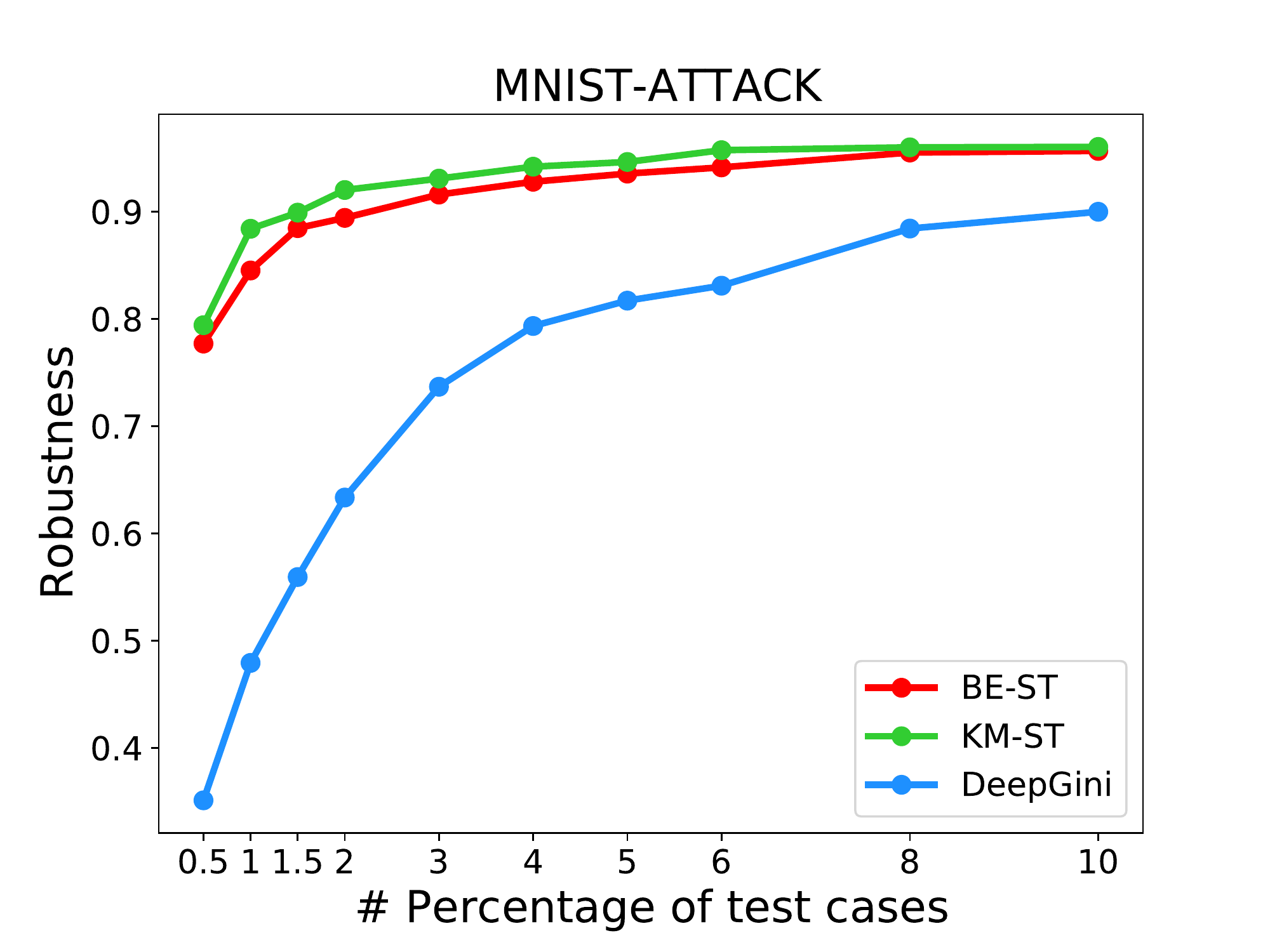}
   \label{fig:ts:mnist}
\end{subfigure}%
\begin{subfigure}[b]{0.25\textwidth}
   \centering 
   \includegraphics[height=1.45in]{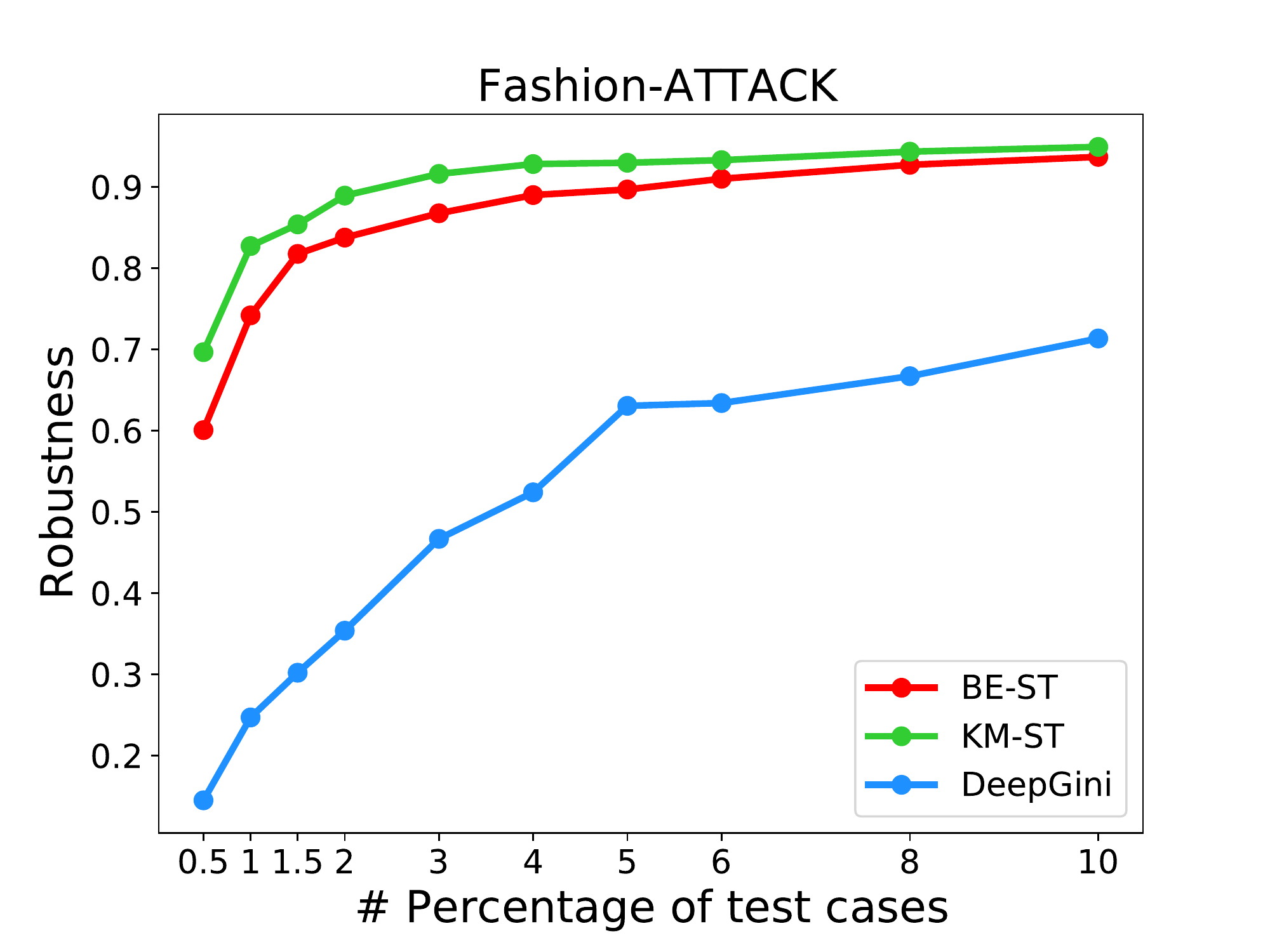}
   \label{fig:ts:fashion}
\end{subfigure}%
\begin{subfigure}[b]{0.25\textwidth}
   \centering 
   \includegraphics[height=1.45in]{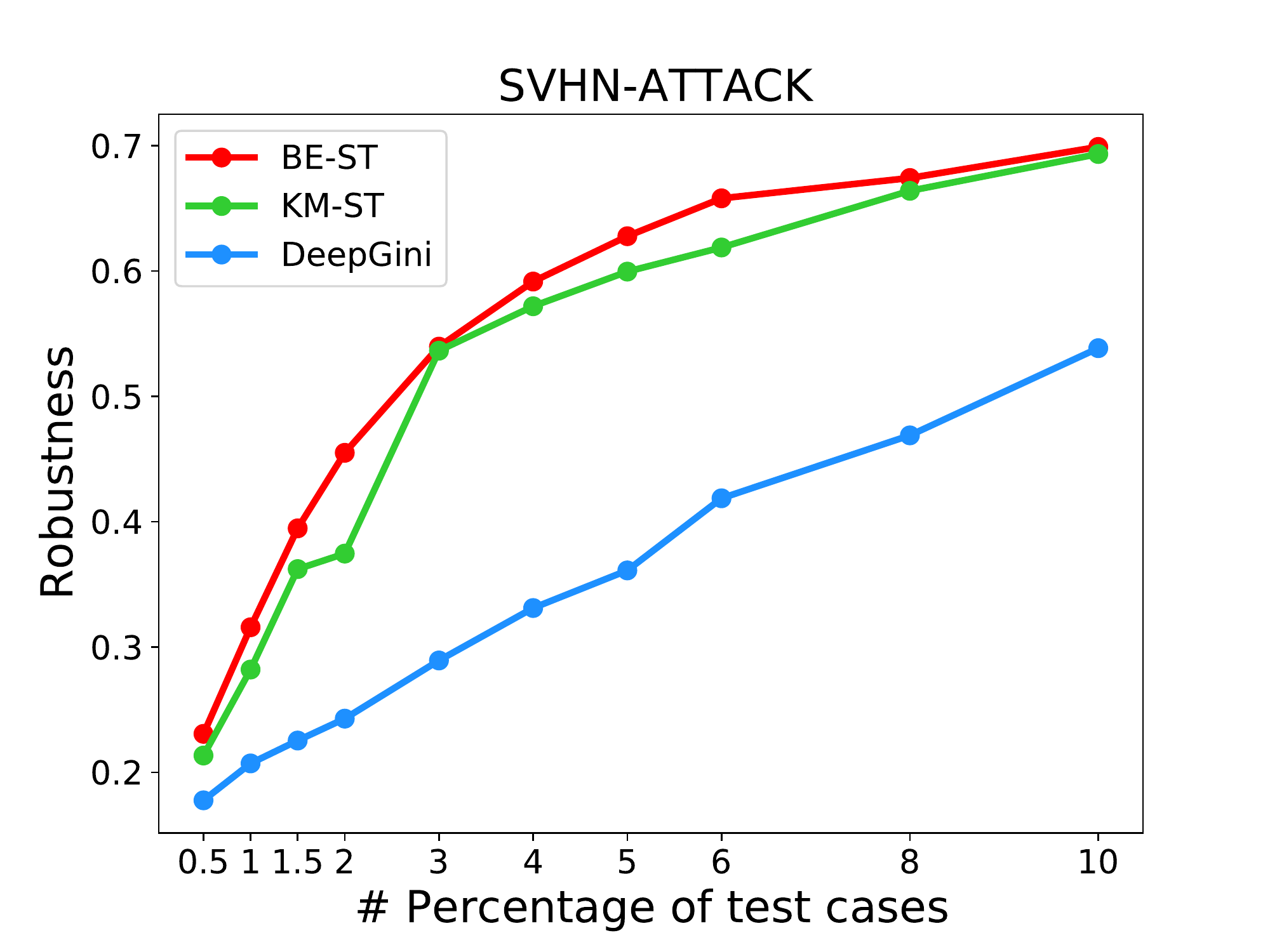}
   \label{fig:ts:svhn}
\end{subfigure}%
\begin{subfigure}[b]{0.25\textwidth}
   \centering 
   \includegraphics[height=1.45in]{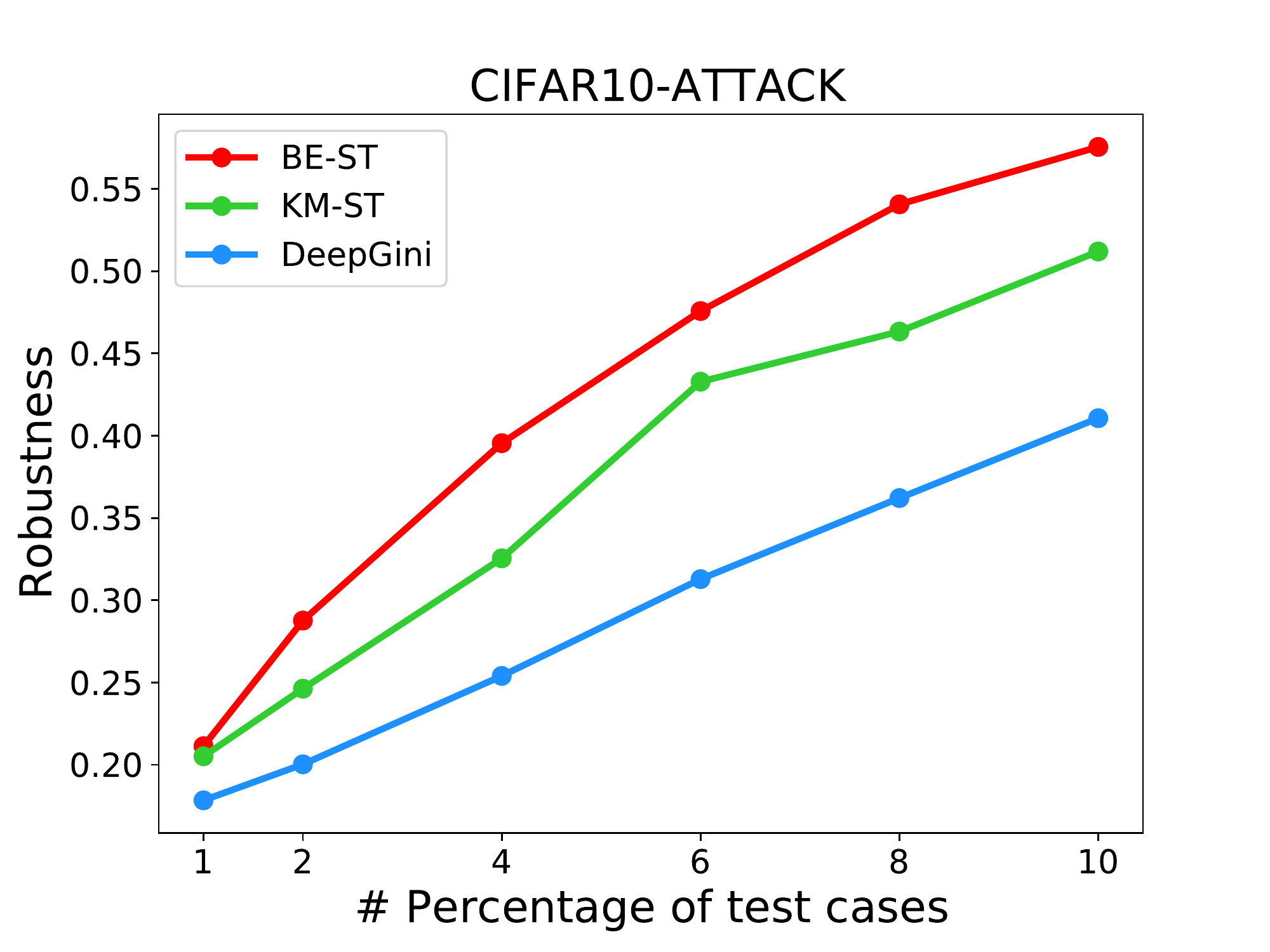}
   \label{fig:ts:cifar}
\end{subfigure}%

\begin{subfigure}[b]{0.25\textwidth}
   \centering 
   \includegraphics[height=1.45in]{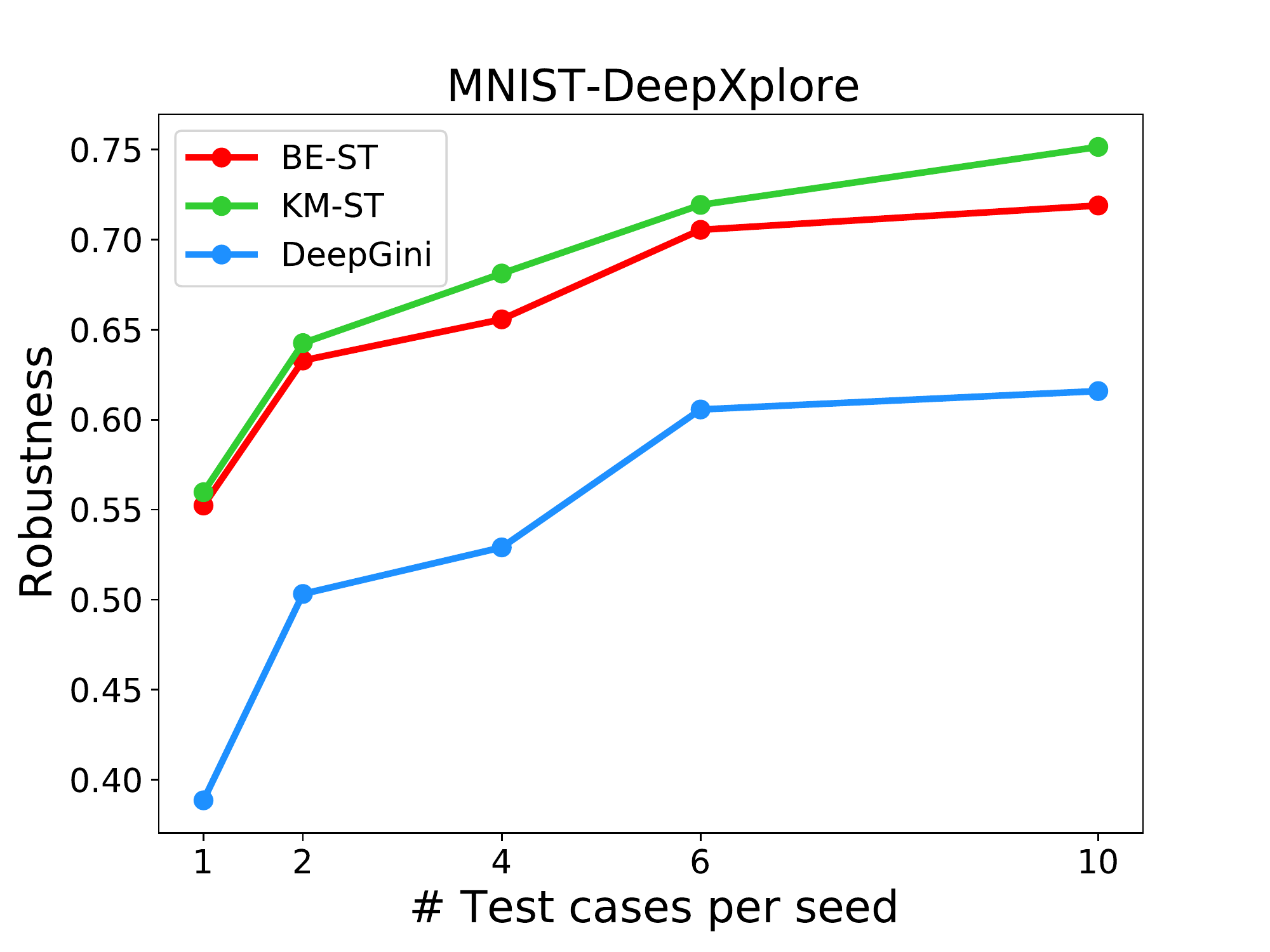}
   \label{fig:ts:mnist}
\end{subfigure}%
\begin{subfigure}[b]{0.25\textwidth}
   \centering 
   \includegraphics[height=1.45in]{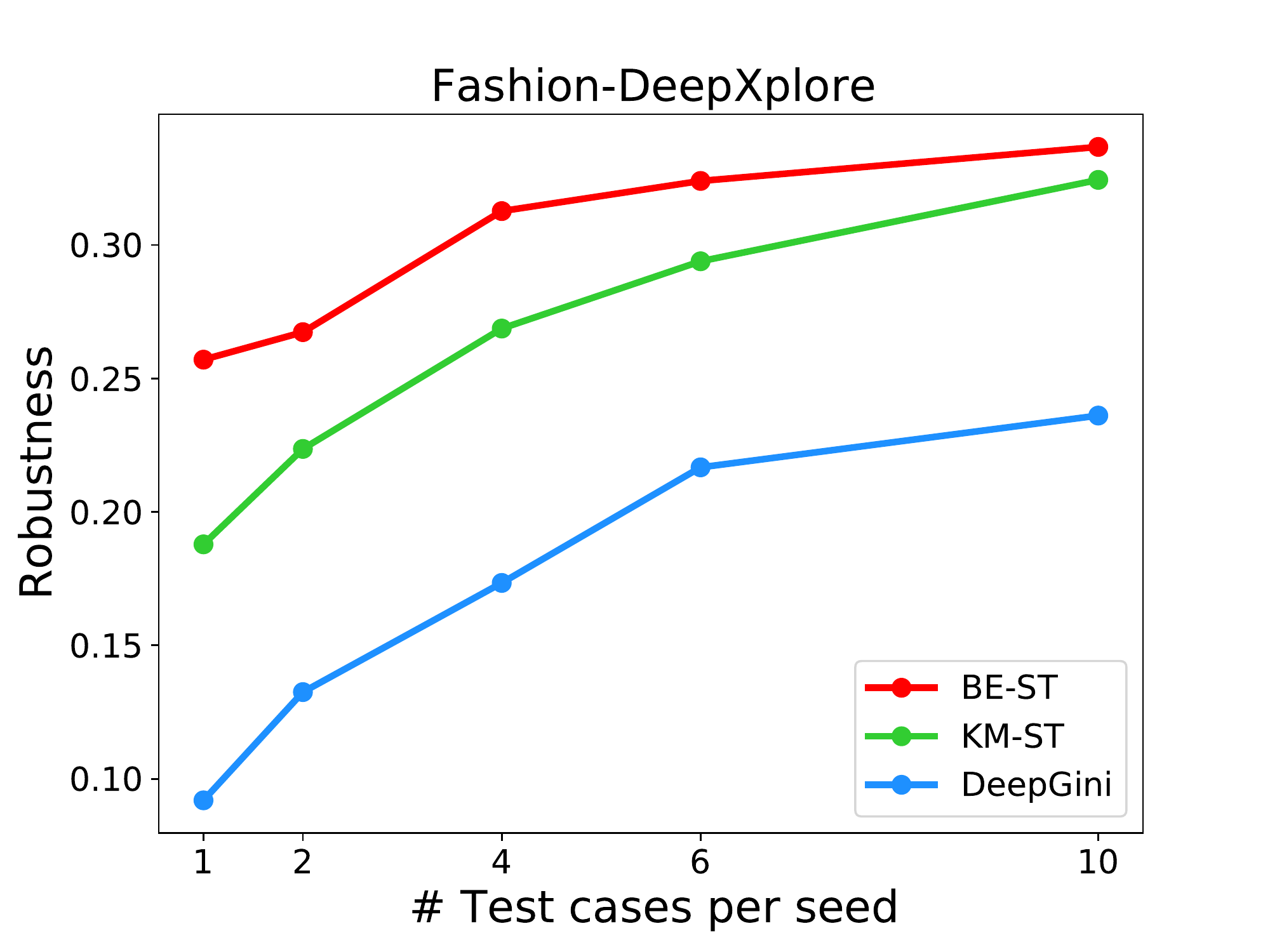}
   \label{fig:ts:fashion}
\end{subfigure}%
\begin{subfigure}[b]{0.25\textwidth}
   \centering 
   \includegraphics[height=1.45in]{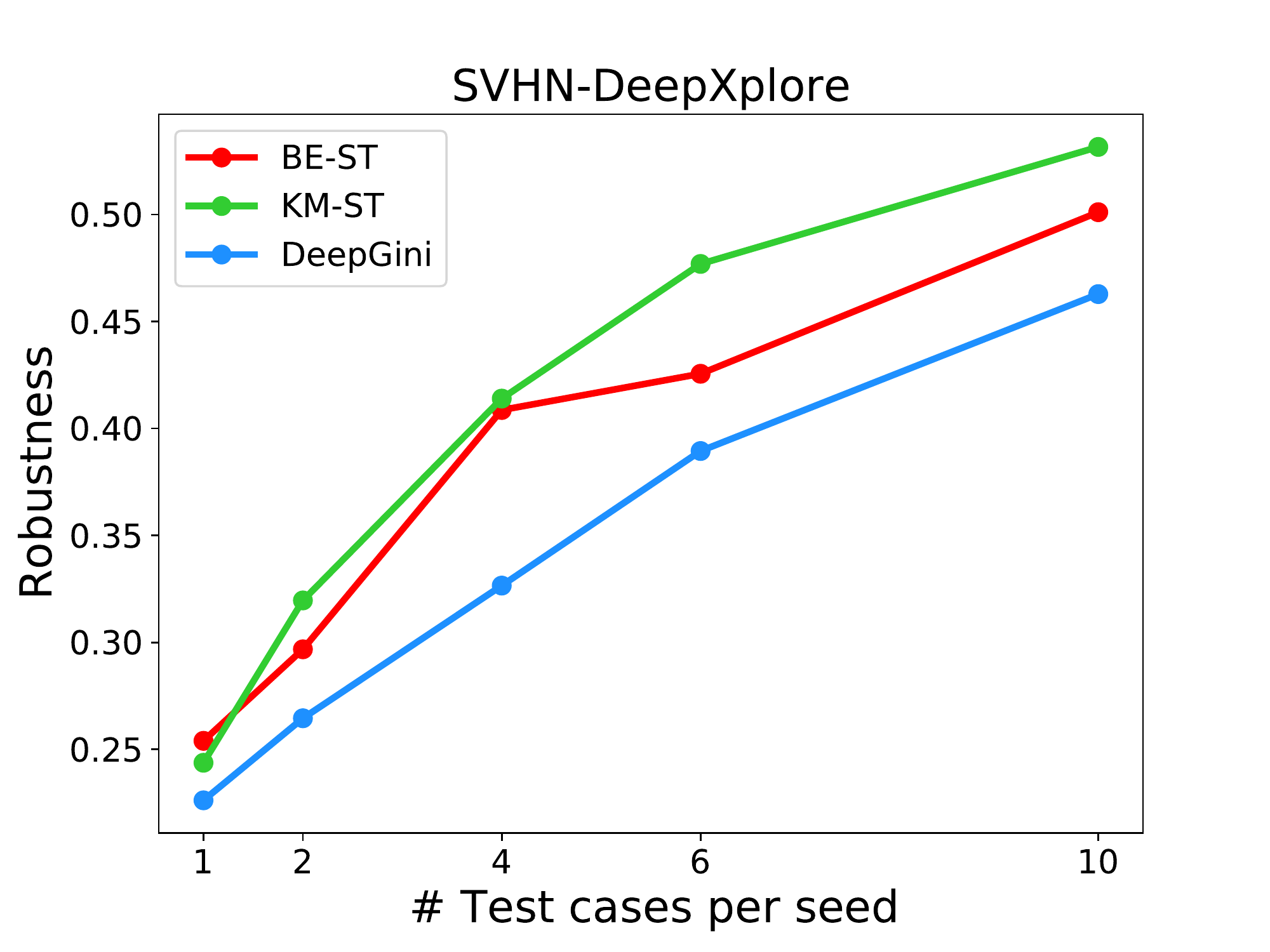}
   \label{fig:ts:svhn}
\end{subfigure}%
\begin{subfigure}[b]{0.25\textwidth}
   \centering 
   \includegraphics[height=1.45in]{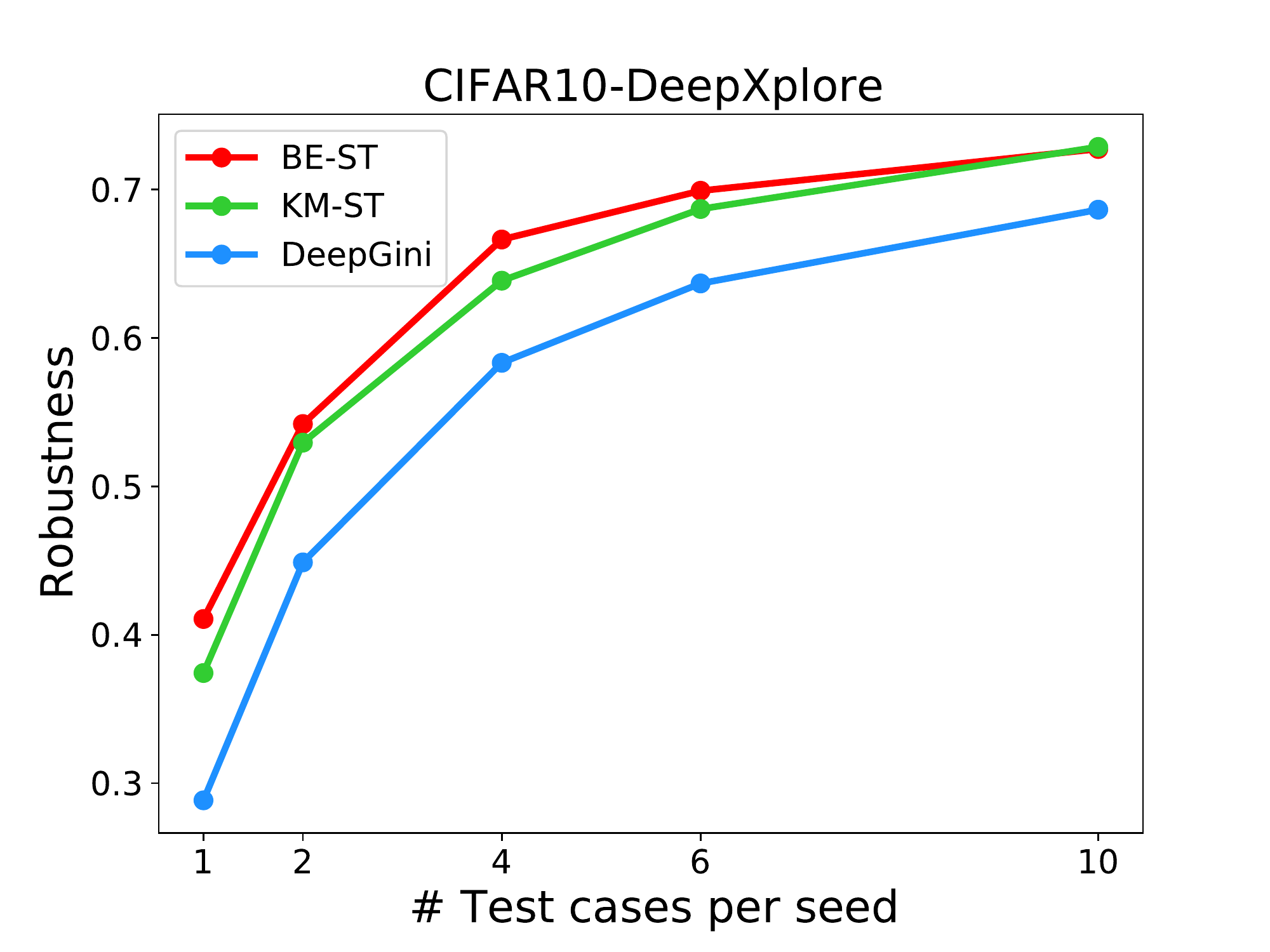}
   \label{fig:ts:cifar}
\end{subfigure}%

\begin{subfigure}[b]{0.25\textwidth}
   \centering 
   \includegraphics[height=1.45in]{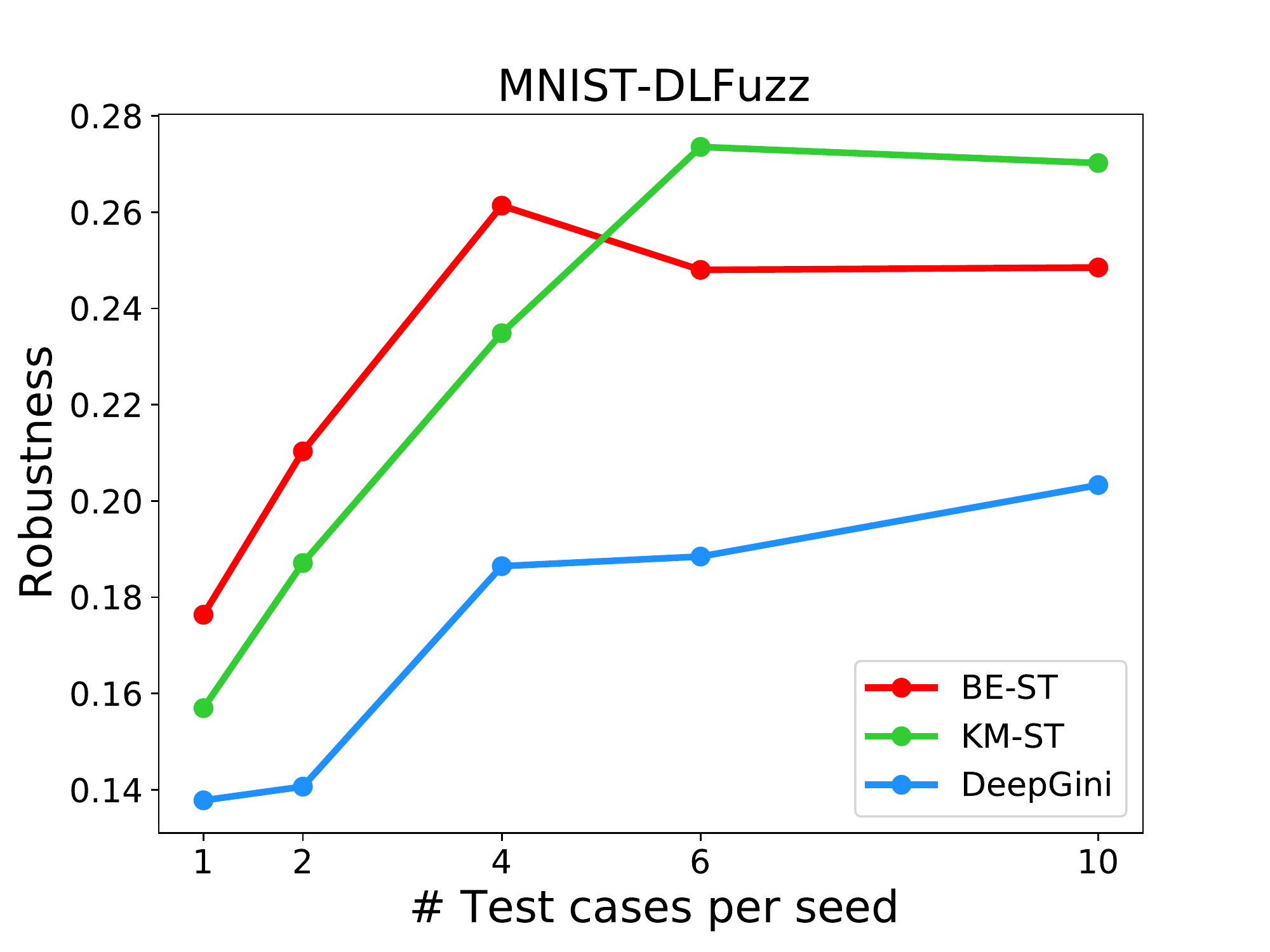}
   \label{fig:ts:mnist}
\end{subfigure}%
\begin{subfigure}[b]{0.25\textwidth}
   \centering 
   \includegraphics[height=1.45in]{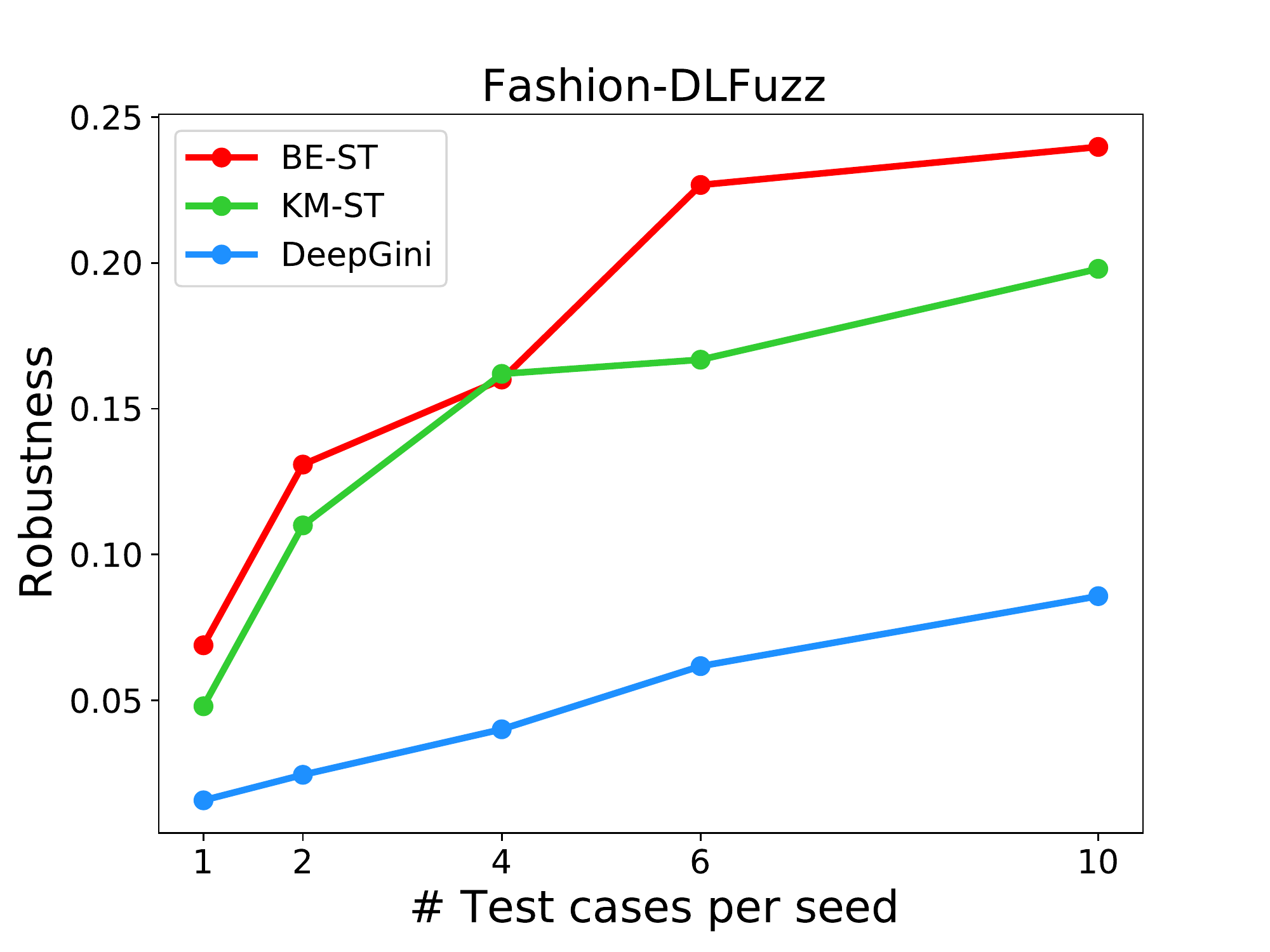}
   \label{fig:ts:fashion}
\end{subfigure}%
\begin{subfigure}[b]{0.25\textwidth}
   \centering 
   \includegraphics[height=1.45in]{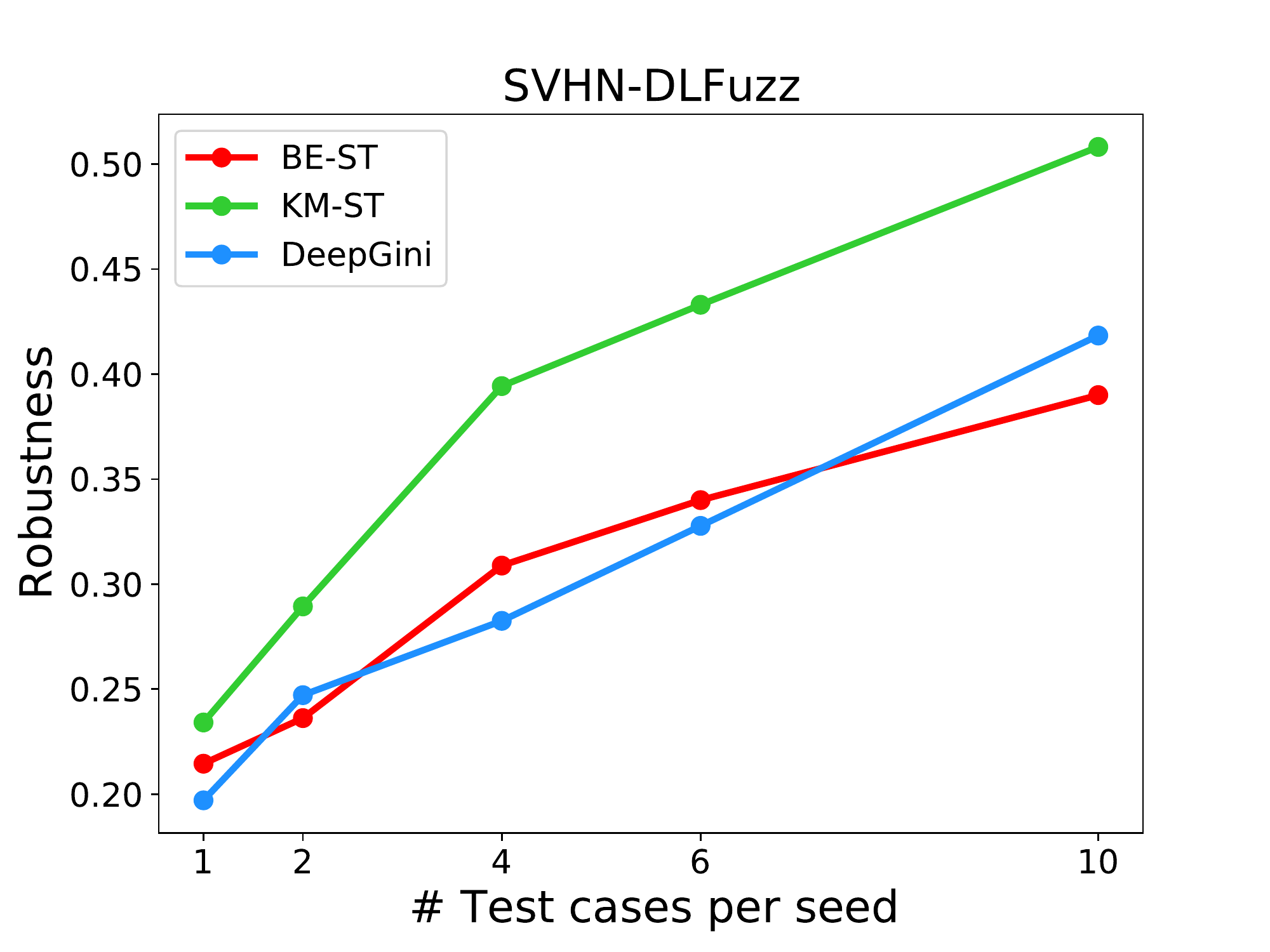}
   \label{fig:ts:svhn}
\end{subfigure}%
\begin{subfigure}[b]{0.25\textwidth}
   \centering 
   \includegraphics[height=1.45in]{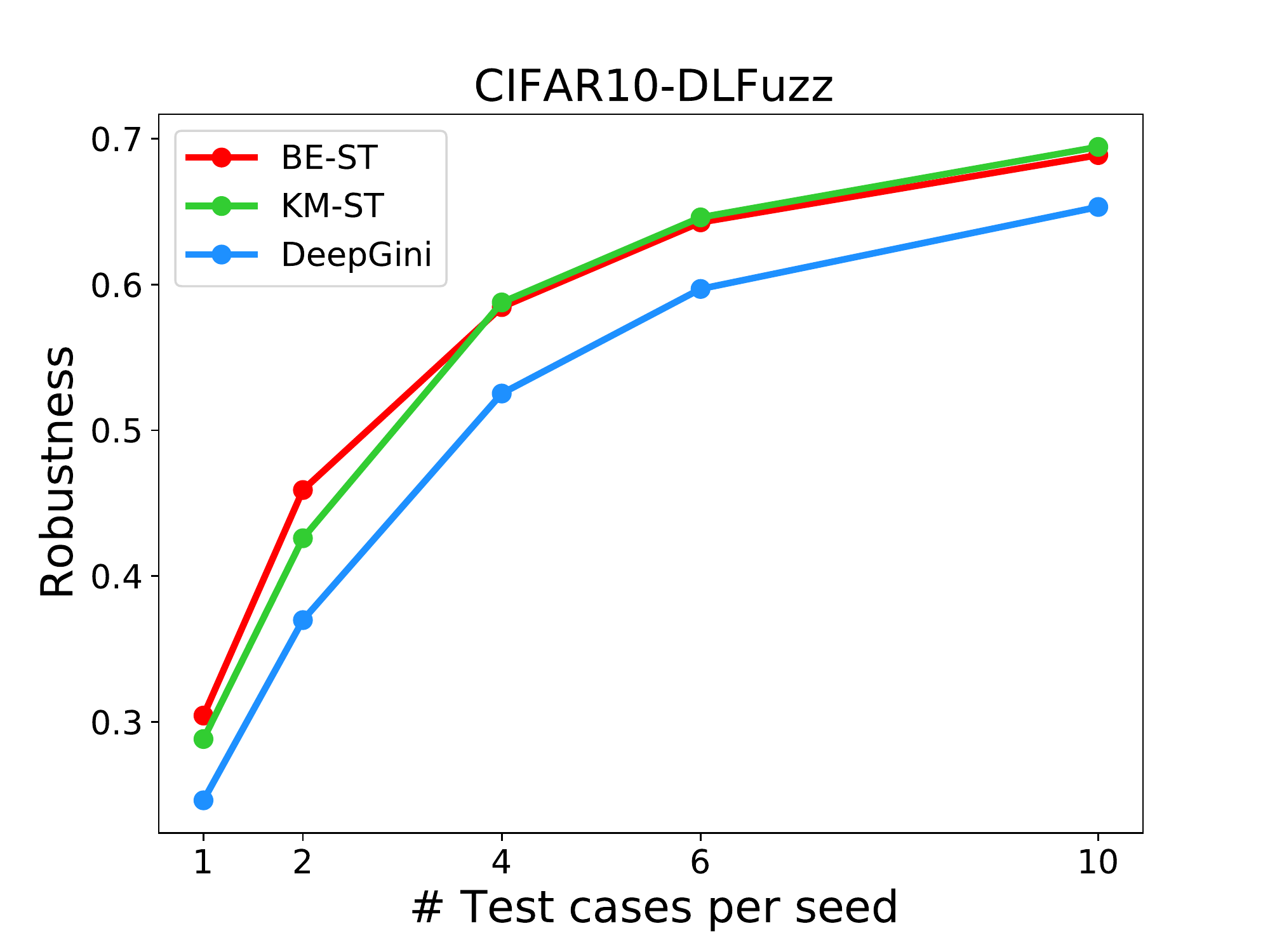}
   \label{fig:ts:cifar}
\end{subfigure}%

\begin{subfigure}[b]{0.25\textwidth}
   \centering 
   \includegraphics[height=1.45in]{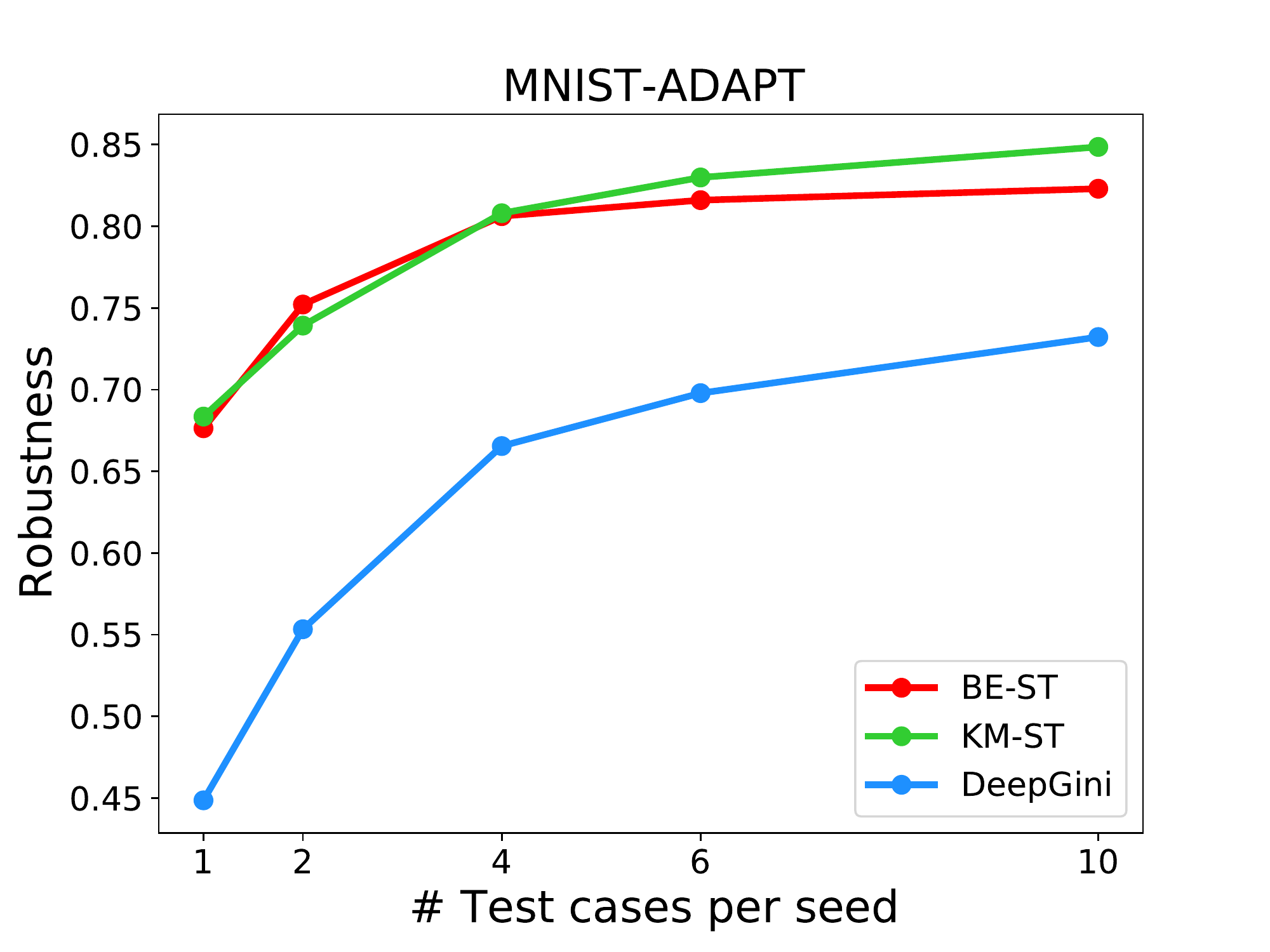}
   \label{fig:ts:mnist}
\end{subfigure}%
\begin{subfigure}[b]{0.25\textwidth}
   \centering 
   \includegraphics[height=1.45in]{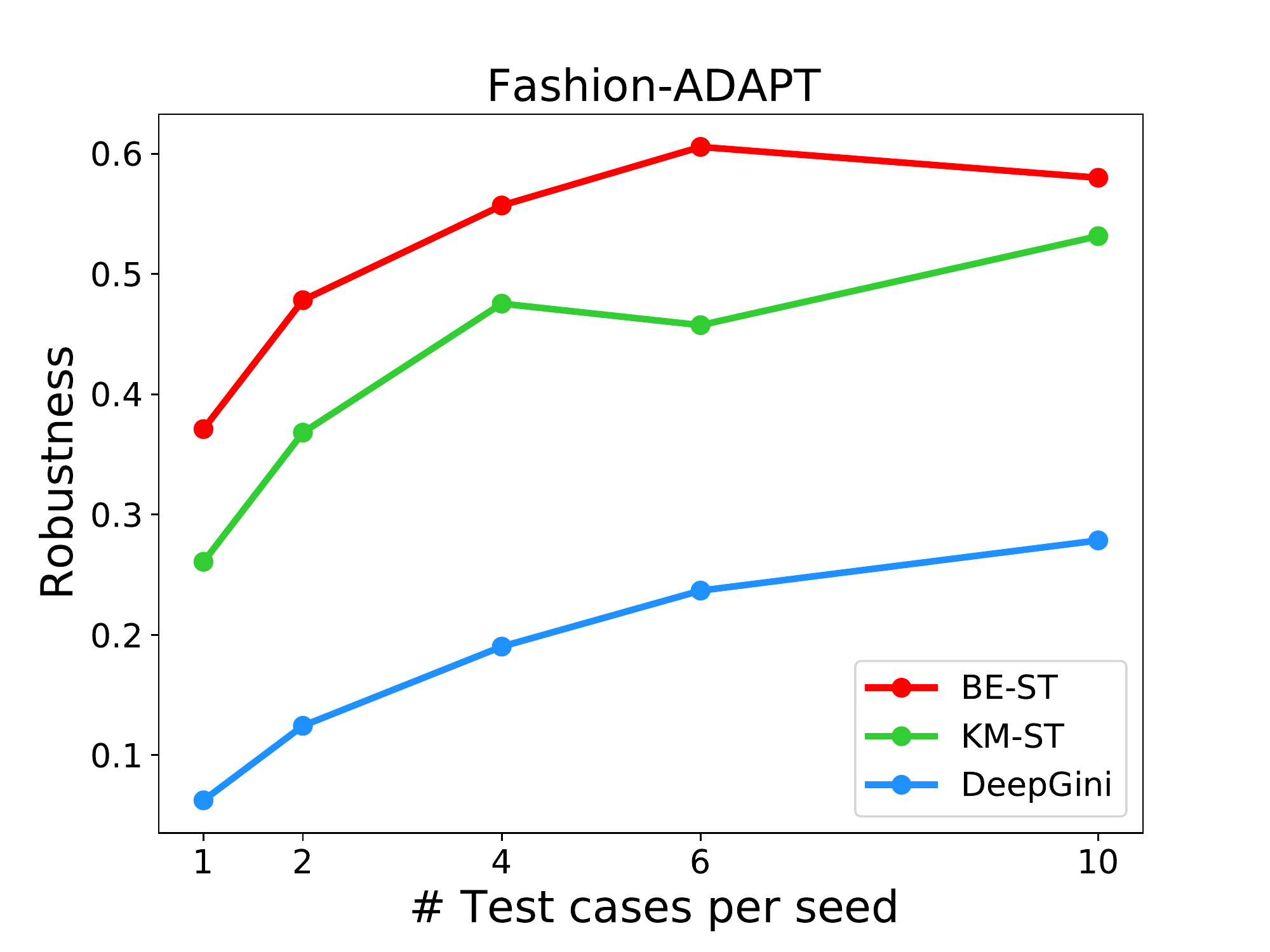}
   \label{fig:ts:fashion}
\end{subfigure}%
\begin{subfigure}[b]{0.25\textwidth}
   \centering 
   \includegraphics[height=1.45in]{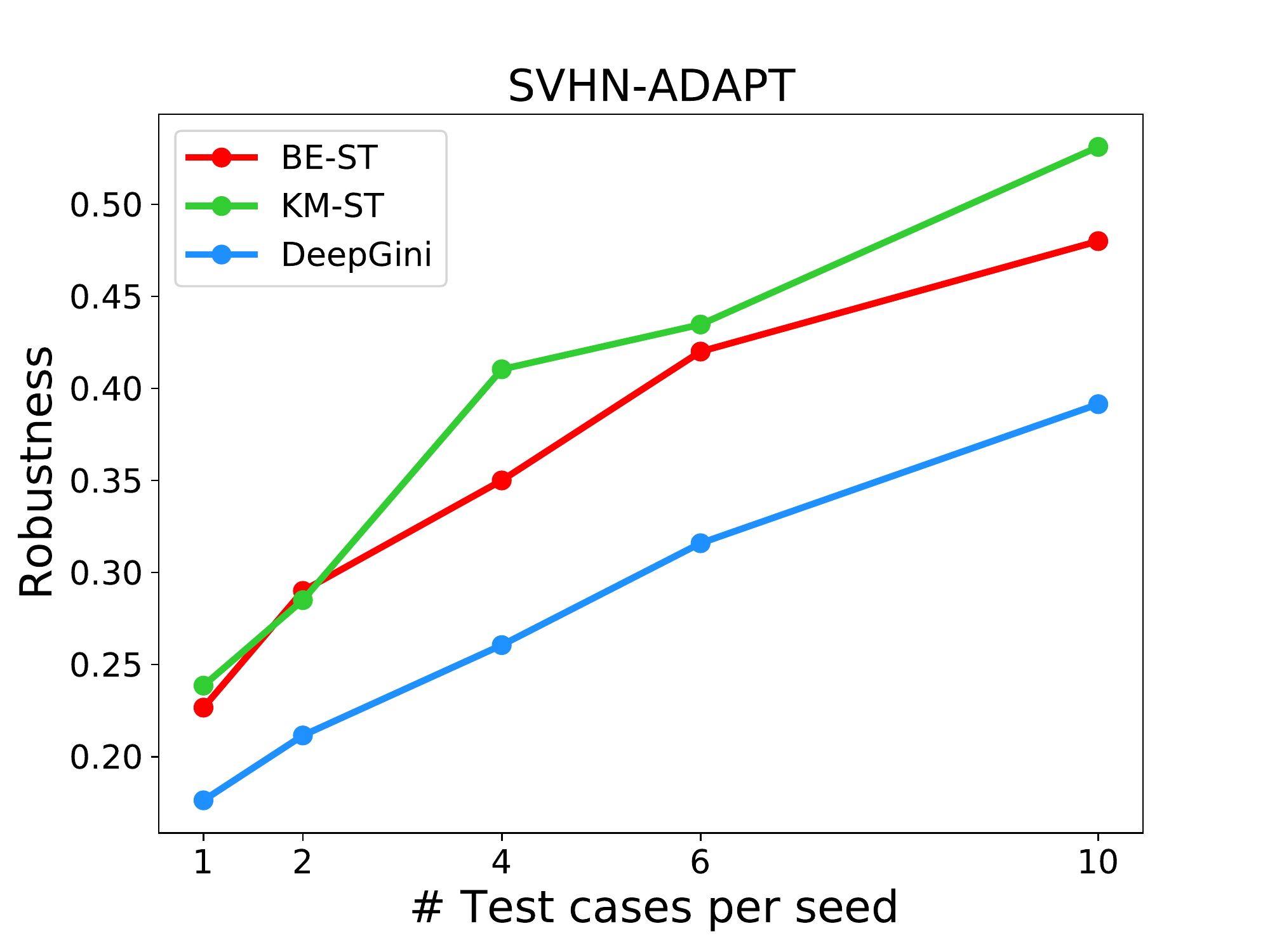}
   \label{fig:ts:svhn}
\end{subfigure}%
\begin{subfigure}[b]{0.25\textwidth}
   \centering 
   \includegraphics[height=1.45in]{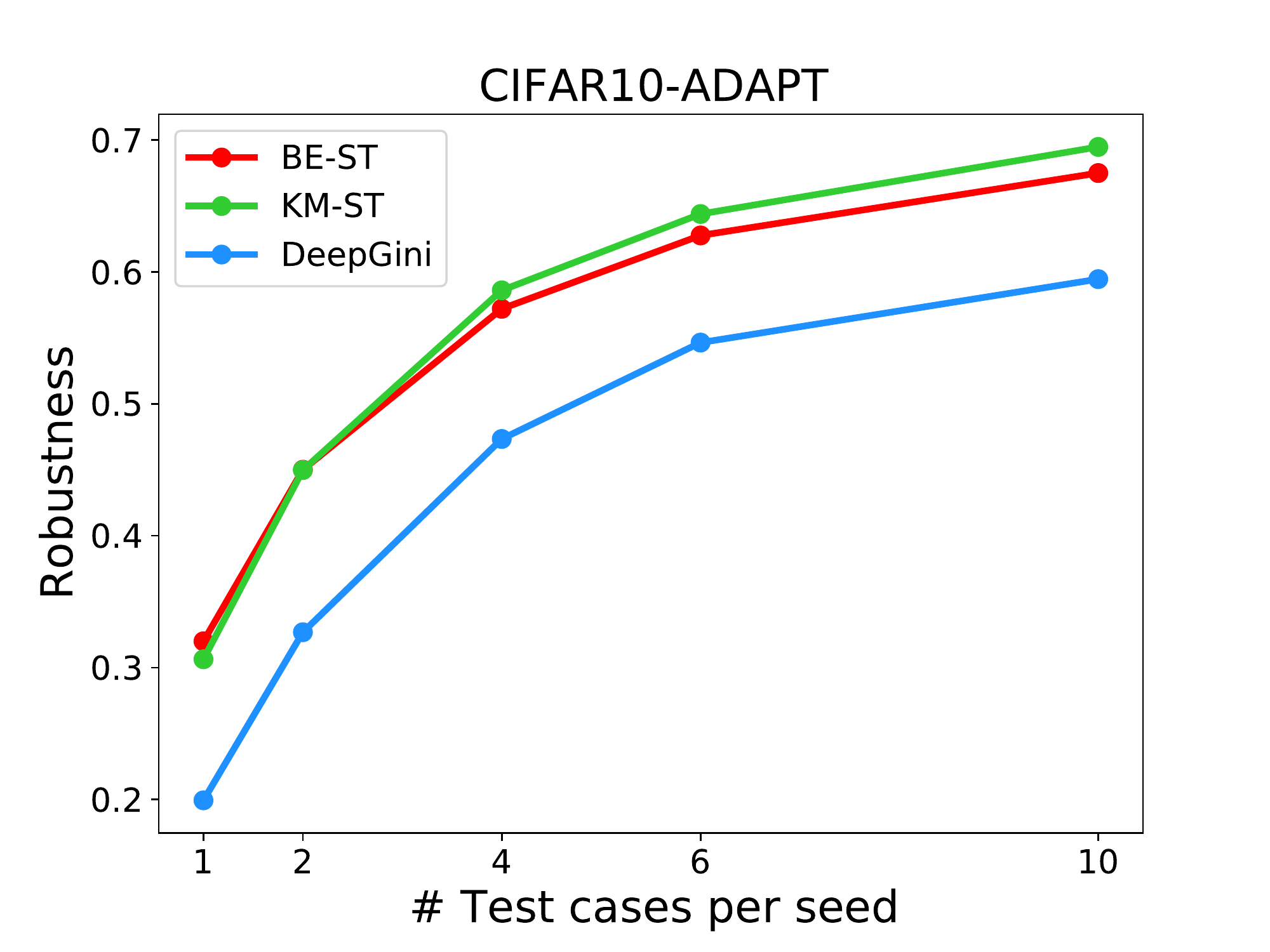}
   \label{fig:ts:cifar}
\end{subfigure}%

   \caption{Test case selection and robustness improvement with different strategies.}
   \label{fig:ts}
\end{figure*}

Fig. \ref{fig:ts} shows the results. We observe that for all the strategies, the retrained model obtained improved resilience to adversarial examples to some extent. Besides, the model's robustness steadily improves as we augment more test cases (from 1\% to 10\%) for retraining. However, we could also observe that in almost all cases (except 1 case), our FOL guided strategies (both BE-ST and KM-ST) have significantly better performance than DeepGini, i.e., achieving 30.48\%, 84.62\%, 54.91\% and 35.92\% more robustness improvement on average for the four different sets of test cases. The reason is that FOL is able to select test cases which have higher and more diverse loss than DeepGini (as shown in Fig. \ref{fig:loss:st} previously), which are better correlated with model robustness. Meanwhile, we observe that the retrained models maintain high accuracy on the test set as well (as summarized in Tab. \ref{tb:acc}).  

Besides, we observe that different test case generation algorithms obtain different robustness improvements. Among DeepXplore, DLFuzz and ADAPT, ADAPT and DLFuzz have the highest (53.39\% on average) and lowest (31.18\% on average) robustness improvement respectively while DeepXplore is in the middle (48.36\% on average). Adversarial attacks often achieve higher robustness improvement than all three neuron coverage-guided fuzzing algorithms for simpler datasets such as MNIST, Fashion-MNIST and SVHN. This casts shadow on the usefulness of the test cases generated by neuron coverage-guided fuzzing algorithms in improving model robustness and is consistent with \cite{limit,misleading,meaningless}. 

We further conduct experiments to evaluate and compare how robust the retrained models are when using adversarial examples generated in different ways: one from the attacks (Fig. \ref{fig:ts}), and the other from different testing algorithms. 
We summarize the result in Tab. \ref{tb:trans}. We observe that the robustness drops noticeably (which is especially the case for CIFAR10), i.e., 18.64\%, 26.41\% and 23.09\% for DeepXplore, DLFuzz, and ADAPT each on average (compared to the results in Fig. \ref{fig:ts}). Nevertheless, our test case selection strategies still outperform DeepGini in all cases. This shows that adversarial examples from adversarial attacks alone are insufficient. It is necessary to improve the diversity of test cases for retraining from a perspective that is well correlated with model robustness.


\begin{table}[]
\centering
\caption{Test accuracy of model before and after retraining with 10 percent of generated test cases using adversarial attacks.}
\begin{tabular}{|l|ll|}
\hline
Dataset       & Original & Retrained \\ \hline
MNIST         & 99.02\%  & 98.95\%   \\
Fashion-MNIST & 90.70\%  & 90.63\%   \\
SVHN          & 88.84\%  & 87.13\%   \\
CIFAR10       & 90.39\%  & 90.13\%   \\
\hline
\end{tabular}
\label{tb:acc}
\end{table}

\begin{table*}[]
\caption{Robustness performance of models (retrained using adversarial examples from attack algorithms)
against test cases generated by DL testing tools.}
\centering
\scalebox{0.85}{
\begin{tabular}{|l|llll|llll|llll|}
\hline
              & DeepXplore &         &          &         & DLFuzz  &         &          &         & ADAPT   &         &          &         \\ \cline{2-13}
Dataset       & BE-ST      & KM-ST   & DeepGini & Average     & BE-ST   & KM-ST   & DeepGini & Average     & BE-ST   & KM-ST   & DeepGini & Average     \\ \hline
MNIST         & 86.12\%    & 80.56\% & 73.74\%  & \textbf{80.14\%} & 76.39\% & 74.73\% & 65.59\%  & \textbf{72.24\%} & 82.60\% & 75.68\% & 70.36\%  & \textbf{76.21\%} \\
Fashion-MNIST & 51.57\%    & 47.97\% & 34.14\%  & \textbf{44.56\%} & 38.15\% & 35.44\% & 27.16\%  & \textbf{33.58\%} & 50.55\% & 47.50\% & 31.92\%  & \textbf{43.32\%} \\
SVHN          & 37.10\%    & 38.29\% & 27.26\%  & \textbf{34.55\%} & 32.83\% & 34.83\% & 25.34\%  & \textbf{31.00\%} & 25.71\% & 28.51\% & 19.15\%  & \textbf{24.46\%} \\
CIFAR10       & 25.25\%    & 20.16\% & 12.92\%  & \textbf{19.44\%} & 18.28\% & 14.20\% & 9.31\%   & \textbf{13.93\%} & 22.37\% & 18.48\% & 12.08\%  & \textbf{17.64\%} \\ \hline

Average       & \textbf{50.01\%}    & \textbf{46.75\%} & \textbf{37.01\%}  &  & \textbf{41.41\%} & \textbf{39.8\%} & \textbf{31.85\%}   &  & \textbf{45.31\%} & \textbf{42.54\%} & \textbf{33.36\%}  &  \\
\hline
\end{tabular}
}
\label{tb:trans}
\end{table*}

\begin{table*}[]
\centering

\caption{Comparison of FOL-fuzz and ADAPT. $a/b$: a is the result of FOL-fuzz and b is the result of ADAPT.}

\begin{tabular}{|l|ll|ll|ll|}
\hline
              & 5 min       &            & 10 min      &            & 20 min      &            \\ \cline{2-7}
Dataset       & \# Test case & Robustness$\uparrow$ & \# Test case & Robustness$\uparrow$ & \# Test case & Robustness$\uparrow$ \\ \hline
MNIST         & 1692/2125        & 33.62\%/18.73\%       & 3472/4521        & 48.04\%/36.46\%       & 7226/8943        & 68.02\%/54.38\%       \\
Fashion-MNIST & 4294/5485        & 40.75\%/6.74\%        & 8906/10433       & 53.88\%/14.94\%       & 18527/21872       & 69.03\%/27.24\%       \\
SVHN          & 6236/8401        & 24.25\%/21.3\%       & 12465/17429       & 30.42\%/27.52\%       & 24864/33692       & 39.99\%/34.51\%    \\
CIFAR10       &      1029/1911     &     18.62\%/17.03\%       &   2006/3722          &     22.07\%/18.12\%       &       4050/6947      &     27.36\%/20.54\%       \\
\hline
Average       &      \textbf{3313/4480}    &  \textbf{29.31\%/15.95\%}          &    \textbf{6712/9026}         &    \textbf{38.6\%/24.26\%}        &       \textbf{13667/17864}      &  \textbf{51.1\%/34.17\%}          \\
\hline
\end{tabular}

\label{tb:fuzz}
\end{table*}

We thus have the following answer to RQ2:
\begin{framed}
\noindent \emph{Answer to RQ2: FOL guided test case selection is able to select more valuable test cases to improve the model robustness by retraining.
}\end{framed}

\vspace{1mm}
\noindent\textbf{RQ3: How effective and efficient is our FOL guided fuzzing algorithm?} To answer the question, we compare our FOL guided fuzzing algorithm (FOL-Fuzz) with state-of-the-art neuron coverage-guided fuzzing algorithm ADAPT as follows. We run FOL-fuzz and ADAPT for a same period of time, (i.e., 5 minutes, 10 minutes and 20 minutes) to generate test cases. Then we retrain the model with the test cases to compare their robustness improvement. 
The hyper-parameters for FOL-Fuzz are set as follows: $\xi=10^{-18},\ k=5,\ \lambda=1,\ iters=3,\ learning\_rate=0.1$. The parameters for ADAPT are consistent with Tab. \ref{tb:tg}.

Tab. \ref{tb:fuzz} shows the results. We could observe that within the same time limit, ADAPT generates slightly more adversarial examples, i.e., 10457 compared to 7897 of FOL-Fuzz. A closer look reveals that ADAPT tends to generate a lot of test cases around a seed towards improving the neuron coverage metrics. However, not all these tests are meaningful to improve model robustness. On the other hand, FOL-Fuzz is able to discover more valuable test cases. We could observe that using FOL-Fuzzed test cases (although less than ADAPT) to retrain the model significantly improves the model's robustness than ADAPT, i.e., 39.67\% compared to 24.79\% of ADAPT on average. 

We thus have the following answer to RQ3:
\begin{framed}
\noindent \emph{Answer to RQ3: FOL-Fuzz is able to efficiently generate more valuable test cases to improve the model robustness.
}\end{framed}

\subsection{Threats to Validity}

First, our experiments are based on a limited set of test subjects in terms of datasets, types of adversarial attacks and neuron coverage-guided test case generation algorithms. Although we included strong adversarial attack like PGD and state-of-the-art coverage-guided generation algorithm ADAPT, it might be interesting to investigate other attacks like C\&W \cite{cw} and JSMA \cite{jsma}, and fuzzing algorithms like DeepHunter \cite{deephunter}. Second, we adopt an empirical approach to evaluate the model robustness which might be different with different kinds of attacks used. So far it is still an open problem as for how to efficiently measure the robustness of DL models. We do not use more rigorous robustness metric like CLEVER \cite{clever} because it is input-specific and has high cost to calculate (e.g., hours for one input).
Third, our testing framework requires a robustness requirement as input which could be application-specific and is relevant to the model as well. In practice, users could 
adjust the requirement dynamically. 





%% file: Related_Work.tex
\section{Related works}
\label{sec:rel}

This work is mainly related to the following lines of works on building more robust deep learning systems.

\paragraph{Deep Learning Testing} Extensive DL testing works are focused on designing testing metrics to expose the vulnerabilities of DL systems including neuron coverage \cite{deepxplore}, multi-granularity neuron coverage \cite{deepgauge}, neuron activation conditions \cite{concolic} and surprise adequacy~\cite{surprise}. Along with the testing metrics, many test case generation algorithms are also proposed including gradient-guided perturbation~\cite{deepxplore,adf}, black-box~\cite{feature_guided_test} and metric-guided fuzzing~\cite{deephunter,adapt,dlfuzz}. However, these testing works lack rigorous evaluation on their usefulness in improving the model robustness (although most of them claim so) and have been shown to be ineffective in multiple recent works \cite{limit,misleading,meaningless}. Multiple metrics have been proposed in the machine learning community to quantify the robustness of DL models as well \cite{xu2012robustness,weng2018towards,brendel2019accurate,clever}. However, most of them are used to evaluate local robustness and hard to calculate. Thus these metrics are not suitable to test directly. Our work bridges the gap by proposing the FOL metric which is strongly correlated with model robustness and integrate retraining into the testing pipeline for better quality assurance.


\paragraph{Adversarial Training} 
The key idea of adversarial training is to improve the robustness of the DL models by considering adversarial examples in the training phase. There are plenty of works on conducting adversarial attacks on DL models (of which we are not able to cover all) to generate adversarial examples such as FGSM ~\cite{fgsm}, PGD \cite{pgd} and C\&W~\cite{cw}.  
Adversarial training in general may overfit to the specific kinds of attacks which generate the adversarial examples for training~\cite{pgd} and thus can not guarantee robustness on new kinds of attacks. Later, robust training \cite{pgd} is proposed to train robust models by solving a saddle point problem described in Sec. \ref{sec:meth}. DL testing complements these works by generating more diverse adversarial examples. 


%% file: Conclusion.tex
\section{Conclusion}
\label{sec:con}

In this work, we propose a novel robustness-oriented testing framework RobOT for deep learning systems towards improving model robustness against adversarial examples. The core of RobOT is a metric called FOL to quantify both the value of each test case in improving model robustness (often via retraining) and the convergence quality of the model robustness improvement. We also propose to utilize the proposed metric to automatically fuzz for more valuable test cases to improve model robustness. 
We implemented RobOT as a self-contained open-source toolkit. Our experiments on multiple benchmark datasets verify the effectiveness and efficiency of RobOT in improving DL model robustness, i.e., with 67.02\% increasement on the adversarial robustness that is 50.65\% higher than the state-of-the-art work DeepGini.